\documentclass[fleqn]{article}

\usepackage{amsmath}
\usepackage{hyperref}
\usepackage[]{graphicx}
\usepackage{color}
\usepackage[utf8]{inputenc}
\usepackage{marginnote}
\usepackage[UKenglish]{babel}
\usepackage{supertabular}
\usepackage[margin=0.75in]{geometry}

\usepackage{mathbbol}
\usepackage{amssymb}
\usepackage{textcomp}
\allowdisplaybreaks[2]

\newcommand{\tpsi}{\widetilde{\psi}}
\newcommand{\tj}{\widetilde{j}}
\newcommand{\tS}{\widetilde{S}}

\newcommand{\tX}{\widetilde{X}}
\newcommand{\hM}{\widehat{M}}
\newcommand{\rb}{{\mathbf{r}}}
\newcommand{\bK}{{\mathbf{K}}}

\newcommand{\bh}{{\mathbf{h}}}
\newcommand{\bp}{{\mathfrak{p}}}
\newcommand{\bx}{{\mathfrak{x}}}
\newcommand{\hx}{\hat{\mathbf{x}}}
\newcommand{\hy}{\hat{\mathbf{y}}}
\newcommand{\hz}{\hat{\mathbf{z}}}
\newcommand{\rmd}{{\rm{d}}}
\newcommand{\rmi}{{\rm{i}}}
\newcommand{\rme}{{\rm{e}}}

\newcommand{\appropto}{\mathrel{\vcenter{
  \offinterlineskip\halign{\hfil$##$\cr
    \propto\cr\noalign{\kern2pt}\sim\cr\noalign{\kern-2pt}}}}}

\hbadness=\maxdimen
\vbadness=\maxdimen
\begin{document}

\title{Three-dimensional model of a split-crystal x-ray and neutron interferometer}
\author{C.\ P.\ Sasso, G.\ Mana, and E.\ Massa\\
INRIM -- Istituto Nazionale di Ricerca Metrologica,\\ strada delle cacce 91, 10135 Torino, Italy}
\date{}%
\maketitle

\begin{abstract}
The observation of neutron interference by using a crystal interferometer having a separate analyser opens the way to the construction and operation of interferometers with vast arm separation and length. Setting the design specifications requires a three-dimensional dynamical-theory model of their operation. In this paper, we develop the needed three-dimensional mathematical framework, which also comprises coherent and incoherent illuminations, and apply it to study the visibility of the interference fringes.
\end{abstract}

\setlength{\mathindent}{0pt}

\section{Introduction}

Since its first demonstrations by Bonse and Hart in 1965 \cite{Bonse:1965b} and Rauch and collaborators in 1974 \cite{Rauch:1974}, perfect-crystal interferometry has been a powerful tool to perform experimental physics and metrology with x-rays and neutrons \cite{Rauch:2000,Klein_2009,Pignol:2015,Massa:2020,Sponar:2021}.

\indent
Because of the short wavelength, the interferometer's crystals require atomic-scale linear and angular alignments and stability. These requirements prompted the consideration of monolithic arrangements, where the splitting and recombining crystals are carved in a single block of a highly perfect silicon crystal. However, a monolith is not compatible with large separations of the interferometer arms, where optics and samples might be inserted, and extended arm-lengths, where weak interactions can be amplified. For instance, when using neutrons having 0.1 nm wavelength, in a 1 m long arm there would be $10^{10}$ waves, so that any minute retardation should accumulates a detectable phase shift.

A solution to these limitations are interferometers consisting of separated crystals. In the case of x-rays, the separation allowed accurate determinations of the silicon lattice parameter \cite{Massa_2011,Massa_2015}. Also, it allowed extending the sample area and volume in phase-contrast imaging \cite{Yoneyama:2002}. Eventually, an interferometer composed of six separate diffracting crystals allowed characterising the temporal coherence of 10 keV pulses from an x-ray free-electron laser \cite{Osaka:2017}. An attempt to operate a split-crystal interferometer with neutrons is reported in \cite{Uebbing:1991}, but it did not succeed in achieving the interference. Neutron interferometry with physically split gratings and cold or very cold neutrons is reported in \cite{Zouw:2000,Pruner:2006,Pushin_2017}.

Recently, Lemmel and collaborators \cite{Lemmel_2021,Massa:2022} operated a neutron interferometer having a split analyser crystal and achieved a proof-of-principle demonstration that split-crystal interferometry with neutrons is possible. This demonstration opens the way to the realisation of skew-symmetric interferometers operating with neutrons and having crystal separations up to the meter scale. In these interferometers, already operated with x-rays \cite{Becker:1974,Yoneyama:2002}, the split crystal embeds also a mirror. They are insensitive to axial misalignments and allow long and spaced interferometer arms, as well as scans of the arm length and Bragg angle alignment.

The object of this paper is to develop a mathematical framework by which the interferometers' operation could be modeled and analysed and design specifications provided. These goals require knowing the effect of three-dimensional misalignments between the crystals splitting and recombining the neutron wave-function \cite{Becker:1974}. Therefore, we extended the dynamical theory of neutron and x-ray diffraction \cite{Bonse:1965a,Bauspiess:1976,Bonse:1977,Vittone:1997a} to model crystal interferometers that operate with both x-rays and neutrons in three dimensions.

Since the paraxial approximations of the (time independent) Schr\"odinger and Helmholtz equations -- relevant to neutron and x-ray propagations, respectively -- are the same, we introduce Fourier-components of the periodic electric susceptibility that mimic also the components of the periodic Fermi pseudo-potential seen by neutrons propagating in perfect crystals. The spatial coherence of the incoming particles (photons or neutrons) limits the interference visibility. Therefore, we studied both coherent and incoherent sources.

In this paper, we consider only a symmetric interferometer where the analyser is free to move with respect to the splitter-mirror pair. In addition to the greater simplicity, since further improvements require a three-dimensional study of the systematic effects, this choice was also driven by the determination of the $^{28}$Si lattice parameter using an x-ray interferometer having a separate analyser \cite{Massa_2011,Massa_2015}. A split-crystal skew-symmetric interferometer can be studied along the same lines.

The paper is organised as follows. After reviewing the neutron and x-ray propagations in free-space, symmetrically-cut crystals, and triple-Laue interferometers, the sections \ref{coherent} and \ref{incoherent} introduce the Gaussian wave-packet and density-matrix used to describe coherent and incoherent x-ray and neutron sources. The coherent and incoherent operation of the interferometer are discussed in sections \ref{s:coherent} and \ref{s:incoherent}.

We describe the crystal fields as the components of a (quanto-mechanical) state vector \cite{Bonse:1977}. This choice allows us to use matrix descriptions of optical components and simplifies the study of the interferometer, the description of which can be built by assembling simple elements. The x-rays and neutrons leaving the interferometer are described by propagating a single-particle Gaussian wave-packet (coherent source case) and the density-matrix of a Gaussian Schell-model of the beam (partially coherent source case) through the interferometer. Eventually, we quantify the effect of the spatial coherence of the source on the visibility of the pendell\"osung, moiré, and travelling fringes observed when rotating and translating the analyser crystal.

To model the propagation in misaligned and displaced crystals, we determine the linear operator that changes the representations of the x-ray and neutron single-particle states from that used to propagate them through the splitter and mirror crystals to that seen by a roto-translated analyser.

All the computations were carried out with the aids of Mathematica \cite{Mathematica}; the relevant notebook is given as supplementary material. To view and interact with it, download the Wolfram Player free of charge \cite{Player}.

\section{Dynamical theory of the interferometer operation}

\subsection{X-ray and neutron states inside a crystal}

We assume symmetrically cut and plane-parallel crystal slabs. The normal $\hz$ to the slab surfaces and the reciprocal vector $\bh=2\pi\hx\big/d$ ($d$ is the spacing of the diffracting planes) defines the reference frame (see Fig. \ref{fig01}). The position vector $\rb=(\bx, z)$ is split in the $\bx=(x, y)$ (lying in the crystal surface, with the $y$ axis pointing up) and $z$ (normal to the crystal surface) components.

Monochromatic x-rays and neutrons inside a crystal behave like a quantum two-level system, a superposition of two independent states spanning a two-dimensional Hilbert space $V_2$. In this case, the basis, labelled as $o$ and $h$, are plane waves satisfying the Bragg condition. Their (complex) amplitudes, $\psi_{o,h}(\bx;z)$, are slowly-varying and square-integrable functions of the transverse coordinates, which are element of the $L_2(\mathbb{R}^2)$ Hilbert space. They propagate along $z$, an optical axis that plays the role of fictitious time.

By using the Dirac bra-ket notation to ease the calculations \cite{Cohen:2019}, we introduce the single-particle state vector \cite{Vittone:1997a,Mana:2004}
\begin{equation}\label{Dket}
 |\psi(z)\rangle = |\psi_o(z)\rangle|o\rangle + |\psi_h(z)\rangle|h\rangle ,
\end{equation}
where, by setting $n=o, h$,
\begin{equation}\label{Doh}\begin{aligned}
  &\langle \bx| \psi(z) \rangle        = | \psi(\bx;z) \rangle = \psi_o(\bx;z)|o\rangle + \psi_h(\bx;z)|h\rangle , \\
  &\langle \bx, n | \psi(z) \rangle    = \psi_n(\bx;z) , \\
  &| o \rangle                         = \rme^{\rmi \bK_o\cdot\rb} \left( \begin{array}{c}
                                                                                1 \\ 0 \\
                                                                              \end{array} \right) , \hspace{3mm}
  | h \rangle                         = \rme^{\rmi \bK_h\cdot\rb} \left( \begin{array}{c}
                                                                                0 \\ 1 \\
                                                                              \end{array} \right) .
\end{aligned}\end{equation}
It belongs to the tensor product $L_2(\mathbb{R}^2) \otimes V_2$ of the $L_2(\mathbb{R}^2)$ space of the square-integrable two-variable functions and the two-dimensional vector space $V_2$. We use the plus sign for the phase of plane waves, the $2\times 1$ matrix representation of $V_2$, and, in the x-ray case and if not otherwise specified, we consider only a polarisation state, parallel or orthogonal to the reflection plane. In addition,
\begin{equation}\label{Koh}
  \bK_{o,h} = K ( \gamma\hz \mp \alpha\hx )
\end{equation}
are the kinematical wave vectors satisfying the Bragg conditions $\bK_h = \bK_o + \bh$, $|\bK_o|=|\bK_h|=K=2\pi/\lambda$, $\gamma=\cos(\Theta_B)$ and $\mp\alpha=\mp\sin(\Theta_B)$ are the direction cosines of $\bK_{o,h}$, $\Theta_B$ is the Bragg angle, and $\lambda$ is x-ray (neutron) the wavelength. We consider a coplanar geometry, that is, $\bK_{o,h}, \bh$, and $\hz$ are in the same (horizontal) reflection-plane.

\subsection{Free-space propagation}

Neglecting gravity, by using the reciprocal-space representation $\langle \bp|\psi(z)\rangle$ (see the appendix \ref{appendix:1}), the free space propagation of x-rays and neutrons, $|\psi(z)\rangle=F(z)|\psi(0)\rangle$, is given by
\begin{equation*}
 \langle \bp |F(z)| \bp' \rangle = F(\bp;z)\delta(\bp-\bp') ,
\end{equation*}
where $\bp=p\hx+q\hy$ is the variable conjugate to $\bx$,
\begin{equation}\label{Free}
   F(\bp;z) = \exp\left( -\frac{\rmi (p^2 + q^2)z}{2K_z} \right)
          \left( \begin{array}{cc}
            \exp\left[ \rmi pz \tan(\Theta_B) \right] & 0 \\
            0 & \exp\left[ -\rmi pz \tan(\Theta_B) \right] \\
          \end{array} \right) ,
\end{equation}
and $K_z = K\gamma$.

In (\ref{Free}), the $\pm pz \tan(\Theta_B)$ phase contribution corresponds to geometric optics. According to it, in the horizontal plane, the $o$ and $h$ states propagate in the $\bK_{o,h}$ directions. Therefore, considering these terms alone, we have $\psi_{o,h}(x,y; z) = \psi_{o,h}(x \pm z\tan(\Theta_B), y; z=0)$, where the plus and minus signs apply the $o$ and $h$ components, respectively.

The $q_z z = (p^2 + q^2)z/(2K_z)$ phase contribution makes $\psi(\bx;z)$ spreading and the wavefront bending. In addition, $q_z = (p^2 + q^2)/(2K_z)$ makes $|\bK_n+(\bp, q_z)|$ approximating $K$, as required by propagation in a vacuum.

The $\gamma=\cos(\Theta_B)$ factor takes the oblique propagation (with respect to the $z$ axis) of the $|o\rangle$ and $|h\rangle$ states into account. Therefore, $z'=z\big/\gamma$ is the propagation distance along the $\bK_{o,h}$ directions.

\subsection{Laue diffraction in a symmetrically cut crystal}

Neglecting gravity, the Laue transmission by a symmetrically cut crystal, $|\psi(z)\rangle = U_0(z)|\psi(0)\rangle$, is given by the scattering matrix \cite{Vittone:1997a,Mana:2004}
\begin{equation*}
 \langle \bp |U_0(z)| \bp' \rangle = U_0(\bp;z)\delta(\bp-\bp') ,
\end{equation*}
where
\begin{subequations}\begin{eqnarray}\label{U0}
 U_0(\bp;z) &= &A(z) \left(
          \begin{array}{cc}
            T(p;z) & R(p;z) \\
            R(p;z) & T(-p;z)
          \end{array} \right)
  \exp\left( -\frac{\rmi (p^2 + q^2)z}{2K_z} \right) \\
  A(z)   &= &\exp\left(\frac{-\mu_0 z}{2\gamma} \right) \exp\left(\frac{\rmi(n_0-1)Kz}{\gamma} \right) \\
  R(p;z) &= &\frac{\rmi \nu \sin(\zeta\sqrt{\eta^2+\nu^2})}{\sqrt{\eta^2+\nu^2}} ,\\
  T(p;z) &= &\cos(\zeta\sqrt{\eta^2+\nu^2}) + \eta R(p;z)\big/\nu .
\end{eqnarray}\end{subequations}
In these equations, $K_z = K\gamma$, $\zeta=\pi z\big/\Delta_e$ is the dimensionless propagation distance, $\eta=\Delta_e \tan(\Theta_B)p\big/\pi$ is the dimensionless resonance error, $\Delta_e = 2\pi\gamma\big/|K\chi_{\pm h}|$ is the pendell\"osung length, and $\nu=\chi_{\pm h}\big/|\chi_{\pm h}|$. A list of the main symbols used is given in the appendix \ref{appendix:6}.

By extending the dynamical theory terminology \cite{Authier_2001}, we call $p$ the resonance error. It encodes the distance of the plane-wave components, $\tpsi_n(\bp;z)\exp(\rmi \bp\cdot\bx)$, of the $\psi_n(\bx;z)$ wave field from the {\it resonance} condition $p/K = 0$. In fact, the reflection and transmission coefficients $R(p;z)$ and $T(p;z)$ do not depend on $q$ and diffraction is two-dimensional: it occurs in a plane, which is defined by the surface normal and $\bh$. Therefore, $p/K = 0$ is associated to the plane waves fulfilling, at the first order, the Bragg condition. This is a consequence of the approximation made when solving the Helmholtz (Schr\"odinger) equation for the x-ray (neutron) propagation in an infinite crystal, which approximation implies $p, q \ll K$ \cite{Mana:2004}.

In (\ref{U0}), the $(p^2 + q^2)z/(2K_z)$ phase recovers the free-space propagation (\ref{Free}) when taking the $\chi_h \rightarrow 0$ limit (this is equivalent to $\Delta_e \rightarrow \infty$, see the supplementary material), which corresponds to neglect the interaction with the crystal lattice. This term was neglected in the first-order approximations made in \cite{Mana:2004}.

The coefficients of the Fourier expansion of the dielectric susceptibility are
\begin{equation*}
 \chi_h = - \frac{4\pi r_e}{K^2 V_{\rm cell}} \sum_j f_j \rme^{-\rmi \bh\cdot\rb_j} ,
\end{equation*}
where the sum is over all the atoms in the unit cell, $r_e$ is the (classical) electron radius, $V_{\rm cell}$ is the cell volume, and $f_j$ the form factor of the $j$-th atom. They are linked by $\chi_h = - \upsilon_h\big/K^2$ to the coefficients of the Fourier expansion of the scaled periodic potential governing the time-independent Schr\"odinger equation for the neutron wave function inside a crystal \cite{Rauch:2000},
\begin{equation*}
 \upsilon_h = \frac{4\pi}{V_{\rm cell}} \sum_j b_j \rme^{-\rmi \bh\cdot\rb_j} ,
\end{equation*}
where $b_j$ is the neutron (coherent) scattering length of the $j$-th atom and $b_{\rm Si} = 4.15071 \times 10^{-6}$ nm \cite{Ioffe_1998,Dianoux:2002}.

A translation $\mathbf{u}$ changes $\chi_h$ to $\chi_h \exp(\rmi\bh\cdot\mathbf{u})$, but propagation is independent of the reference-frame origin, which we chosen so that $\chi_{+h}=\chi_{-h}$. In addition, since $\exp(\pm\rmi\pi)=-1$, propagation is independent of the $\chi_{\pm h}$ and $\nu$ signs. The $\chi_0$ sign depends on the sign in the exponent of the plane wave functions.

\begin{figure}\centering
\includegraphics[width=0.45\columnwidth]{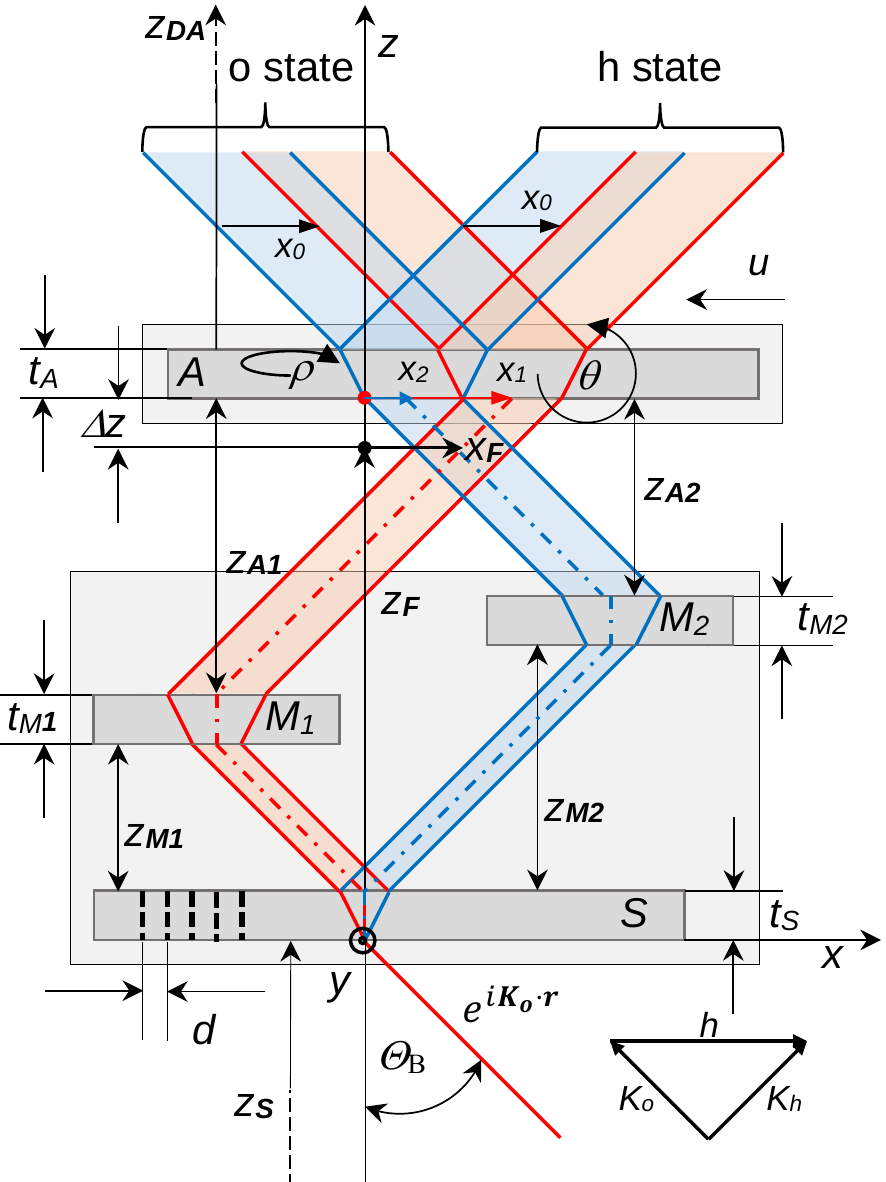}
\includegraphics[width=0.45\columnwidth]{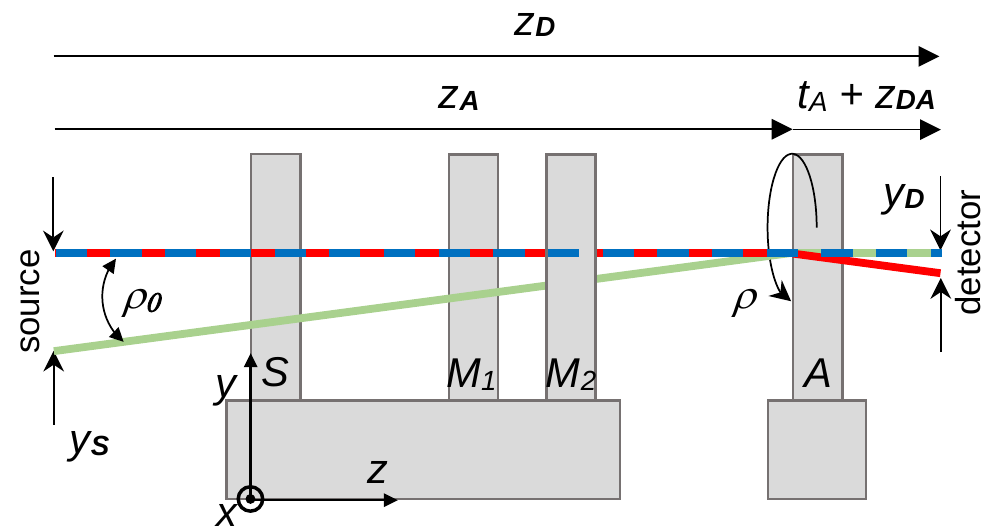}
\caption{Top and side views of a triple-Laue interferometer with a split analyser crystal, which is misaligned by a tilt $\rho$ about the $z$ axis. S splitter, M$_1$ and M$_2$ mirrors, A analyser. The $x$ axis is orthogonal to the diffracting planes. Right. Red and blue rays indicate arms 1 and 2, respectively. $\Theta_B$ is the Bragg angle, $\exp(\rmi\bK_o\cdot\rb)$ is the input $o$ wave, $\Delta z$ is the defocus. $\rho$, $\theta$, and $u$ are the analyser pitch (the rotation about the $z$ axis) and yaw (the Bragg rotation about the vertical $y$ axis) angles and displacement along the $x$ axis, respectively (the arrows indicate the positive rotations and displacement). Left. The red and blue rays are associated to the $q/K=0$ plane-wave components of the single-particle wave function. These rays leave the source collinearly, travel the arms 1 (red) and 2 (blue), interfere in the $o$ state, and have a $y_D=2\rho (t_A+z_{DA}) \tan(\Theta_B)$ offset at the detector, see section \ref{moire-fringes}. The green ray leaves the source at the $\rho_0=2\rho\sin(\Theta_B)$ angle and $y_S = \rho_o z_A\big/\cos(\Theta_B)$ distance, see section \ref{moire-fringes}. It is associated to the $q=h\rho$ components of the single-particle wave function, travels the 1-st arm, and interferes collinearly with the $q/K=0$ components (blue). $z_A$, $t_A+z_{DA}$, and $z_D$ are the source-to-analyser, analyser-to-detector, and source-to-detector distances.}\label{fig01}
\end{figure}

\subsection{Triple Laue interferometer}

Let us consider an interferometer having the analysing crystal separated from the splitter-mirror pair, see Fig.\ \ref{fig01}. The representation of the particle state (\ref{Doh}) uses the kinematical plane waves defined in (\ref{Koh}) as the $|o\rangle$ and $|h\rangle$ states. Since the analyser crystal might be differently oriented and displaced to the splitter-mirror block, this representation cannot be simultaneously used for both. Therefore, we need to change the (laboratory) state leaving the first crystal, $|\psi^L(z)\rangle$, to that seen by a misaligned analyser, $|\psi^A(z)\rangle$. The linear operator $\hM(z)$ that carries out the $|\psi^A(z)\rangle = \hM(z) |\psi^L(z)\rangle$ transformation is given in the appendix \ref{appendix:3}. Hence, the propagation through the interferometer is given by
\begin{subequations}\begin{equation}\label{X}
 |\psi_{\rm out}\rangle = X|\psi_{\rm in}\rangle ,
\end{equation}
where
\begin{equation}
  X = F(z_{DA})\hM^{-1}(\Delta z+t_A)U_0(t_A)\hM(\Delta z)
                  \bigg[ F(z_{A1})P_h U_0(t_{M1})F(z_{M1})P_o
                  + F(z_{A2})P_o U_0(t_{M2})F(z_{M2})P_h \bigg] U_0(t_S) F(z_S) ,
\end{equation}\end{subequations}
and we introduced the projectors $P_n = |n\rangle\langle n|$, $n=o, h$. The symbol meanings are given in Fig.\ \ref{fig01}. The effect of the $\hM^{-1}(\Delta z+t_A)U_0(t_A)\hM(\Delta z)$ transformation is shown in the appendix \ref{appendix:3}.

\subsection{Coherent source.}\label{coherent}

Only one particle (photon or neutron) is inside the interferometer at any given time which, therefore, supposes interference of single particles. We start describing each particle of the incoming beam by the same Gaussian wave packet, $\langle \bx|\psi_{\rm in}\rangle =\langle x|\psi_{\rm in}\rangle \langle y|\psi_{\rm in}\rangle$, monochromatic, separable, originating in $(\bx_0,-z_S)$, and propagating at the $\arctan(\bp_0\big/K)$ angle to the $\bK_o$ direction. Hence, at the source, at a $z_S$ distance from the splitter,
\begin{equation}\label{Do}
 \langle \bx|\psi_{\rm in}\rangle = \psi_{\rm in}(\bx) |o\rangle \propto \exp\left( -\frac{|\bx-\bx_0|^2}{l_0^2} +\rmi\bp_0\cdot\bx \right)\, |o\rangle ,
\end{equation}
where $l_0$ is the radius, which we assumed the same for both the $x$ and $y$ factors; the extension to an elliptical wave packet is trivial. A summary of the equations for the oblique propagation of a Gaussian wave-packet is given in the appendix \ref{appendix:2}. Here and in the following, the proportionality sign indicates that, to avoid non-essential algebraic burdens, we omit any normalisation factor.

The reciprocal-space representations of the $|\psi_{\rm in}\rangle$ (see the appendix \ref{appendix:1} and supplemental material) is
\begin{equation}\label{tDo}
 \langle \bp|\psi_{\rm in}\rangle = \tpsi_{\rm in}(\bp) |o\rangle \propto \exp\left( -\frac{|\bp-\bp_0|^2l_0^2}{4} -\rmi\bp\cdot\bx_0 \right) |o\rangle .
\end{equation}

\subsection{Partially coherent source.}\label{incoherent}

Owing to the limited spatial coherence of x-ray and neutron sources \cite{Rauch:1996,Pushin:2008}, that is, the limited capacity to prepare every particle always in the same state, we consider each incoming particle in a probabilistic superposition of the (separable) single-particle Gaussian states
\begin{equation}\label{spwf}
 \langle \bx|\psi_{\rm in}(x_0,\phi_0,p_0)\rangle \propto \exp\left( -\frac{|\bx-\bx_0|^2}{l_0^2} +\rmi\phi_0 +\rmi\bp_0\cdot\bx \right) |o\rangle .
\end{equation}

The probability density that the Gaussian state $|\psi_{\rm in}(x_0,\phi_0,\bp_0)\rangle$ is centered in $\bx_0$ and has $\phi_0$ phase and $\arctan(\bp_0\big/K)$ propagation angle  is
\begin{equation*}
  p(\bx_0,\phi_0,\bp_0) \propto \exp\left( -\frac{2|\bx_0|^2}{w_0^2} - \frac{\phi_0^2}{2\sigma_\phi^2} - \frac{|\bp_0|^2}{2\sigma_p^2} \right) .
\end{equation*}
Hence, the $\bx_0$, $\phi_0$, and $\bp_0$ are uncorrelated normal variables having zero mean and $w_0^2\big/2$, $\sigma_\phi^2$, and $\sigma_p^2$ variances, where $w_0 \gg l_0$. Without loss of generality, we made the $x_0$ mean and reference frame origins to coincide. Also here we assumed circular profiles for both the single-particle states and their superposition. The extension to elliptical profiles gives no problems.

This mixed state is characterised by the density matrix \cite{Feynman:1998,Cohen:2019}
\begin{equation}\label{jin0}
 j_{\rm in} = \mathbb{E}\big( |\psi_{\rm in}(\bx_a,\phi_a,\bp_a)\rangle\langle \psi_{\rm in}(\bx_b,\phi_b,\bp_b)| \big) ,
\end{equation}
where $\mathbb{E}(.)$ indicates the ensemble average. In the limit as $\sigma_\phi\rightarrow\infty$ and $\sigma_p \gg 1/w_0$, the direct- and reciprocal-space representations of the density matrix are (see the supplementary material)
\begin{subequations}\label{jin1}\begin{eqnarray}
 \langle \bx_1 |j_{\rm in}| \bx_2 \rangle &= &\left( \begin{array}{cc}
        j_{\rm in}(\bx_1,\bx_2)    & 0 \\
        0                          & 0 \\
      \end{array} \right) , \\
 \langle \bp_1 |j_{\rm in}| \bp_2 \rangle &= &\left( \begin{array}{cc}
        \tj_{\rm in}(\bp_1,\bp_2)    & 0 \\
        0                            & 0 \\
      \end{array} \right) \label{jin2},
\end{eqnarray}\end{subequations}
where
\begin{subequations}\begin{eqnarray}\label{Jx12}
 j_{\rm in}(\bx_1,\bx_2)\hspace{-3mm} &\propto\hspace{-3mm} &\exp\left( -\frac{|\bx_1|^2 + |\bx_2|^2}{w_0^2} - \frac{|\bx_1-\bx_2|^2}{2\ell_0^2}
 \right) , \\ \label{Jp12}
 \tj_{\rm in}(\bp_1,\bp_2)\hspace{-3mm} &\propto\hspace{-3mm} &\exp\left( -\frac{2(|\bp_1|^2 + |\bp_2|^2)\ell_0^2 + |\bp_1-\bp_2|^2 w_0^2}{8}
 \right) ,  \hspace{10mm}
\end{eqnarray}\end{subequations}
are the mutual intensities of a Gaussian Schell-model of the particle beam \cite{Schell_1967,Gase:1994},
\begin{equation}
  \ell_0^2 = \frac{l_0^2}{1+l_0^2\sigma_p^2} \label{lc} ,
\end{equation}
and we neglected terms proportional to $l_0/w_0 \ll 1$.

It is easy to check that the density matrix (\ref{jin0}) is positive definite, Hermitian (actually, in this case, symmetric), and has unit trace (or equal to the particle number, depending on the chosen normalisation). Its diagonal elements are the particle densities in the chosen basis. The off-diagonal elements give information about the interferences between the relevant states, i.e., they represent the coherence of the superimposed states.

A diagonal density matrix with equal elements on the diagonal represents a completely incoherent superposition. The density matrix associated to the pure state (\ref{Do}) and (\ref{tDo}) can be obtained by taking the limits as $\ell_0/w_0 \rightarrow\infty$ of (\ref{Jx12}) and (\ref{Jp12}).

\subsubsection{First order correlations.}

The normalized first order correlations (see the supplementary material),
\begin{equation*}
 g_{\rm in}^{(1)}(\bx_1,\bx_2)= \exp\left[-\frac{|\bx_1-\bx_2|^2}{2\ell_0^2} \right] ,
\end{equation*}
highlight that $\ell_0$ measures the (transverse) coherence lengths.

\subsubsection{Free-space propagation.}

The free-space propagation of the density matrix is given by $j_z = F j_{\rm in} F^\dag$ \cite{Friberg:1982}, where the dagger indicates the adjoint and $F$ and $j_{\rm in}$ are given by (\ref{Free}) and (\ref{jin0}), respectively. To exemplify, let us consider the
\begin{equation*}
 \tj_z(q_1,q_2) = \rme^{ \rmi q_1^2 z\big/(2K_z) } \tj_{\rm in}(q_1,q_2) \rme^{ -\rmi q_2^2 z\big/(2K_z) }
\end{equation*}
factor. After transforming it back to the direct space, we obtain
\begin{equation*}
 j_z(y_1,y_2) \propto \exp\left[  -\frac{y_1^2 + y_2^2}{w_z^2} - \frac{(y_1-y_2)^2}{2\ell_z^2} - \frac{\rmi K(y_1^2-y_2^2)}{2r_z} \right] ,
\end{equation*}
where (see the supplementary material)
\begin{subequations}\begin{eqnarray}
 w_z      &= &w_0 \sqrt{ 1 + \frac{4z^2}{K_z^2w_0^2\ell_0^2} }    = \frac{w_0 \ell_z}{\ell_0}, \label{wcz} \\
 \ell_z   &= &\ell_0 \sqrt{ 1 + \frac{4z^2}{K_z^2w_0^2\ell_0^2} } = \frac{\ell_0 w_z}{w_0} , \label{lcz} \\
 r_z      &= &\frac{K_z^2w_0^2\ell_0^2 + 4z^2}{4z} ,
\end{eqnarray}\end{subequations}
are the radius, spatial coherence, and radius of curvature at a distance $z$ from the beam source, respectively.

The coherence length spreads like the radius of a coherent beam having $\ell_0$ source radius and $\theta_\ell = \arctan[2/(K_zw_0)]$ divergence, see (\ref{lcz}). Therefore, propagation increases the beam coherence, which is the content of the van Cittert-Zernike theorem. When $\ell_0/w_0 \rightarrow 0$, that is, when the beam's particles are completely incoherent, as the particles propagate, the coherence length increases as $2z \big/(K_zw_0)$

\subsubsection{Particle density.}\label{particle-density}

The particle density of the propagated Gaussian Schell model,
\begin{equation*}
 S(\bx) = j_z(\bx,\bx) \propto \exp\big(-2|\bx|^2\big/w_z^2 \big) ,
\end{equation*}
behaves like the particle density of a coherent beam having $w_0$ source radius and $\theta_w = \arctan[2/(K_z\ell_0)]$ divergence. When the beam's particles are completely incoherent, $\ell_0/w_0 \rightarrow 0$ case, then $K w_z\rightarrow\infty$. Therefore, a finite divergence is possible only if the source coherence is not null.

\subsection{Interference signal: coherent source}\label{s:coherent}

When all the incoming particles are in the same state $\langle \bp|\psi_{\rm in}\rangle = \tpsi_{\rm in}(\bp)|o\rangle$, they leave the interferometer as
\begin{equation*}
 \langle \bp|\psi_{\rm out}\rangle =  \exp\left( -\frac{\rmi(p^2+q^2)z_D}{2K_z} \right) \sum_{\substack{ {n=o,h} \\ {i=1,2} }} A(t_i) \tpsi_{ni}(\bp) |n\rangle ,
\end{equation*}
where $t_i=t_S+t_{Mi}+t_A$ is the total crystal thickness along the $i$-th arm and, by the application of (\ref{X}) (see the supplementary material),
\begin{subequations}\label{oh12}\begin{eqnarray}
    \tpsi_{o1}(\bp)\hspace{-2mm} &= \hspace{-2mm}&R(p+\theta K_z-\rho q;t_A)  R(p;t_{M1}) T(p;t_S) \tpsi_{\rm in}(p,q-h\rho)
 \exp\left[ -\rmi [ px_1 - qy_S + h(u +\theta\Delta z) ] \right] \label{oh12a},\\
    \tpsi_{o2}(\bp)\hspace{-2mm} &= \hspace{-2mm}&T(p+\theta K_z-\rho q;t_A)  R(p;t_{M2}) R(p;t_S) \tpsi_{\rm in}(p,q)
 \exp( -\rmi px_2 ) ,\\
    \tpsi_{h1}(\bp)\hspace{-2mm} &= \hspace{-2mm}&T(-p-\theta K_z+\rho q;t_A) R(p;t_{M1}) T(p;t_S) \tpsi_{\rm in}(p,q)
 \exp( -\rmi px_1 ) ,\\
    \tpsi_{h2}(\bp)\hspace{-2mm} &= \hspace{-2mm}&R(p+\theta K_z-\rho q;t_A)  R(p;t_{M2}) R(p;t_S) \tpsi_{\rm in}(p,q+h\rho)
 \exp\left[ -\rmi [ px_2 + qy_S - h(u +\theta\Delta z) ] \right] \label{oh12b}.
\end{eqnarray}\end{subequations}
\noindent
In the equations (\ref{oh12}), see Fig.\ \ref{fig01}, $\rho$ and $\theta$ are the analyser rotations about the $z$ and $y$ axis, respectively,
\begin{equation*}
  x_i = \pm (z_{Ai} - z_{Mi})\tan(\Theta_B) ,
\end{equation*}
are the horizontal offsets from the origin of the waves leaving the interferometer (the plus and minus signs refer to the $i= 1, 2$ arms, respectively),
\begin{equation*}
  y_S = 2\rho z_A \tan(\Theta_B) ,
\end{equation*}
is the vertical offset at the source between the plane-wave components interfering collinearly, see section \ref{moire-fringes},
\begin{equation*}
  \Delta z = (z_{A1}-z_{M1}) - (z_{M2}-z_{A2}) ,
\end{equation*}
is the defocus, $z_A = z_S+t_S+z_{Mi}+t_{Mi}+z_{Ai}$ and $z_D=z_A+t_A+z_{DA}$ are the source-to-analyser and source-to-detector distances, respectively, and $K_z=K\gamma$.

Free-space propagation leads to the spatial separation of the $o$ and $h$ states leaving the interferometer into two spatially localised states, $[\tpsi_{o1}(\bx)+ \tpsi_{o2}(\bx)]|o\rangle$ and $[\tpsi_{h1}(\bx)+ \tpsi_{h2}(\bx)]|h\rangle$, whose $i=1, 2$ components overlap and interfere. In (\ref{oh12}), unessential phases shared by the interfering $o$ and $h$ states and second order terms have been omitted.

The equations (\ref{oh12}) chain together the particles' reflections and transmissions along their way through the interferometer. The phases proportional to $p$ and $q$ take their horizontal and vertical shifts into account. Eventually, as the analyser moves, the $h(u+\theta\Delta z)$ phase gives rise to travelling fringes, the period of which is equal to the diffracting plane spacing. This is the foundation of the measurement of the Si lattice parameter and atomic-scale length metrology by x-ray interferometry.

It is worth noting that $2\Delta z=(x_1-x_2)/\tan(\Theta_B)$ and $h=2K\sin(\Theta_B)$. Therefore, the phase term $h\theta\Delta z$ in (\ref{oh12a}) and (\ref{oh12b}) can be rewritten as $h\theta\Delta z = K_z (x_1-x_2)\theta$. This formula shows that the $h\theta\Delta z$ phase originates in the difference $\theta(x_1-x_2)$ of the lengths of the particles' paths through the interferometer.

\begin{figure}\centering
\includegraphics[width=0.45\columnwidth]{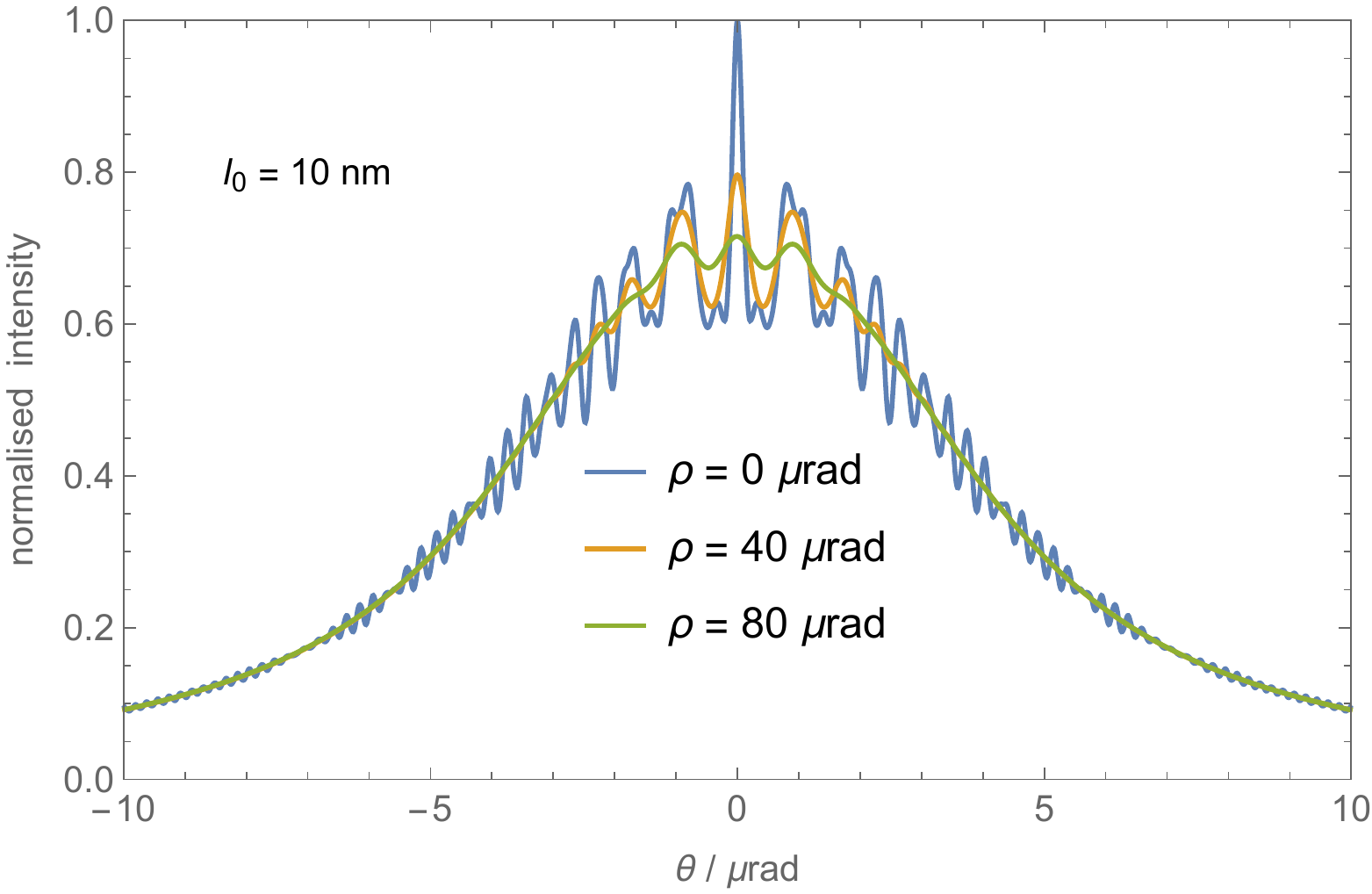}
\includegraphics[width=0.45\columnwidth]{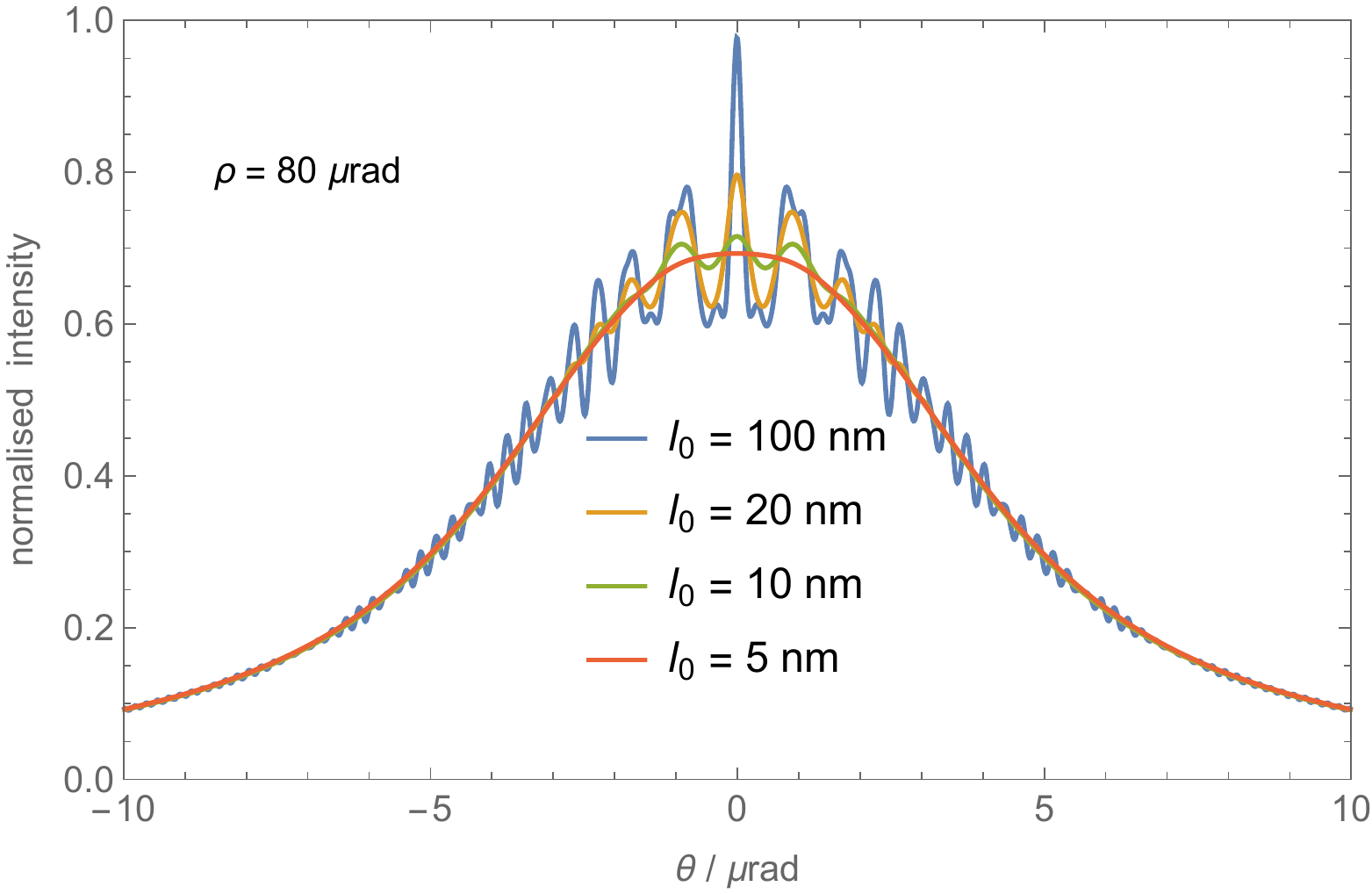}
\caption{Triple Laue rocking curves of a crystal neutron-interferometer having a split analyser, see (\ref{RRR}) and table \ref{parameters}. The abscissa is the Bragg misalignment of the analyser to the splitter-mirror pair. $l_0$ is the (vertical) source size of the single-particle wave-function, see (\ref{Do}), $\rho$ is the pitch misalignment of the analyser.}\label{fig03}
\end{figure}\vspace{-1mm}

\subsubsection{Triple Laue rocking curve.}

Figure \ref{fig03} shows the triple Laue rocking curves \cite{Rauch:1983,Petrascheck:1984},
\begin{equation}\label{RRR}
 I_{\rm RRR}(\theta,\rho) \propto \int_{-\infty}^{+\infty} \big|
 R(p+\theta K_z-\rho q;t_A)  R(p;t_{M2}) R(p;t_S)  \big|^2 \big| \tpsi_{\rm in}(p,q+h\rho) \big|^2\, \rmd p\, \rmd q ,
\end{equation}
of a neutron interferometer having different analyser pitch-angles and illuminated by Gaussian wave functions having different source radii. The simulation parameters are given in table \ref{parameters}; $\tpsi_{\rm in}(p,q)$ is given by (\ref{tDo}), where $p_0/K=q_0/K=0$ and both the horizontal and vertical radii are equal to $l_0$. The crystal thicknesses were chosen so that -- when $\eta = 0$, see Eqs.\ (5c) and (5d) -- the mirror reflectivity is maximum and the reflectivity and transmissibility of the splitter-analyser pair are equal. The $I_{\rm RRR}(\theta)$ symmetry about the $\theta=0$ rad axis originates from $R(p;z)$ and $\tpsi_{\rm in}(p,q)$ -- because of the $p_0/K=0$ choice -- being even functions of $p$.

The $\rho q$ offsets of the null resonance-error, see the argument of $R(p+\theta K_z-\rho q;t_A)$ in (\ref{oh12}-d), shifts the analyser's Bragg alignment, that, otherwise, occurs at $\theta=0$ rad. It makes the Bragg alignment dependent on $q$, i.e., on the out of reflection-plane propagation of the plane-wave components of the particle wave-function.

The averaging of out-of-phase pendell\"osung fringes -- generated by the plane waves propagating at different vertical angles to the reflection plane -- jeopardises their visibility. For any given spread of the (vertical) momentum $q$, see (\ref{tDo}), the pitch angle reduces the fringe visibility, which visibility can be recovered by improving the vertical collimation. In the $Kl_0\rightarrow\infty$, i.e., a cylindrical wave-function having a perfect vertical collimation (approximated in Fig.\ \ref{fig03} by the $l_0 = 100$ nm case), the rocking curve is insensitive to the pitch misalignment.

\begin{table}\hspace{1mm}
\caption{\label{parameters} Parameters used in the numerical simulations.}
\begin{tabular}{ll}
$\chi_0 = \chi_h = -2.382 \times 10^{-6}$    &$\nu = -1$ \\
$n_0 = 1-1.191 \times 10^{-6}$               &$\mu_0 = 0$ \\
$\lambda = 0.190$ nm                         &$d = 192$ pm \\
$K = 33.1$ rad/nm                            &$h = 32.7$ rad/nm \\
$\Delta_e = 69.3$ $\mu$m                     &$\Theta_B = 0.518$ rad \\
$t_S = t_A = 15.75\Delta_e \approx 1.092$ mm &$t_{M1} = t_{M2} = 15.5\Delta_e \approx 1.074$ mm
\end{tabular}
\end{table}

\subsubsection{Defocus.}\label{defocus}

As shown in Fig.\ \ref{fig01}, the interfering waves $\psi_{n1}(\bx)$ are sheared by $x_0 = x_1-x_2 = 2\Delta z \tan(\Theta_B)$ with respect to $\psi_{n2}(\bx)$. We recall that $x_i$ is the horizontal distance from the origin of the beam leaving the interferometer after travelling the $i$-th arm.

This shear is reflected in the appearance of the extra phases $p x_i$ in $\tpsi_{ni}(\bp)$, see the equations (\ref{oh12}). In fact, transforming, for instance, $\tpsi_{o1}(\bp)$ and $\tpsi_{o2}(\bp)$ back to the direct space, and recalling the "time-shifting" property of the Fourier transform, we obtain $\psi_{o1}(x,y) = \psi_{o2}(x-x_0,y)$.

With a plane wave illumination, the offset between $\psi_{n1}(\bx)|n\rangle$ and $\psi_{n2}(\bx)|n\rangle$ develops only an unessential phase difference. Contrary, with a point source the interfering wavefronts are curved and their offset recombines plane-wave components that left the source at different angles and yields a pattern of vertical fringes. In this respect, the interferometer operation is related to the wavefront shearing in light optics. A detailed geometric-optics study is in \cite{Bonse:1971}.

The first-order effect of defocus on the phase of the travelling fringes -- encoded by the $px_i$ phases in the waves (\ref{oh12}) leaving the interferometer -- and, consequently, on the lattice parameter measurement was investigated analytically and experimentally in \cite{Vittone:1997b,Sasso:2021a,Sasso:2021b}.

The present analysis brings into light an additional phase term, $h\theta\Delta z$, associated with the simultaneous existence of the analyser defocus and misalignment. In the split-analyser case here considered, the travelling fringes are visible only if $\Delta z/\Delta_e \approx 0$ \cite{Vittone:1997b}. Therefore, it is a second-order effect.

\subsubsection{Moiré fringes.}\label{moire-fringes}
The analyser pitch angle $\rho$ tilts the reflected waves by $\rho_0 = h\rho\big/K = 2\rho\sin(\Theta_B)$. As shown in Fig.\ \ref{fig01}, the $q$'s plane-wave components of the single-particle wave function superimposing collinearly after travelling the interferometer have an $y_S = \rho_0 z_A\big/\cos(\Theta_B)$ offset at their start. This offset stems from the $qy_S$ phase of $\tpsi_{o1}(\bp)$ and $\tpsi_{h2}(\bp)$ in the same way as discussed in section \ref{defocus}.

For example, from the equations (\ref{oh12}) (where the $q\rho$ offset of the analyser Bragg alignment is neglected), the virtual wave $\tpsi'_{\rm in}(\bp)=\tpsi_{\rm in}(p,q-h\rho)\exp(\rmi qy_S)$ overlaps collinearly $\tpsi_{o2}(\bp)$ when leaving the interferometer. Transforming $\tpsi'_{\rm in}(\bp)$ back to the direct space, we obtain $\psi'_{\rm in}(x,y)=\psi_{\rm in}(x,y-y_S)\exp\big[+\rmi h\rho(y-y_S)\big]$, whose origin and propagation direction are changed by $y_S$ and $\rho_0$, respectively.

The misalignment of the interfering waves yields a pattern of horizontal fringes. Let us neglect the $q\rho$ offset of the analyser Bragg alignment. Hence, considering only the $q$ factor and the $o$ state, the interfering waves are
\begin{equation}\begin{aligned}\label{Do12}
 &\tpsi_{o1}^{\rm out}(q) \propto \tpsi_{\rm in}(q -h\rho) \exp\left( -\frac{\rmi q^2z_D}{2K_z} +\rmi qy_S -\rmi h\rho\Delta y \right) , \\
 &\tpsi_{o2}^{\rm out}(q) \propto \tpsi_{\rm in}(q) \exp\left( -\frac{ \rmi q^2z_D}{2K_z} \right) ,
\end{aligned}\end{equation}
where $y_S = 2\rho z_A \tan(\Theta_B)$ and $z_A$ and $z_D$ are the source-to-analyser and source-to-detector distances, respectively. Also, we assumed a rotation axis vertically displaced by $\Delta y$ from the origin. Hence, the analyser is displaced by $\rho\Delta y$ and the $h\rho\Delta y$ term appears in the $\tpsi_{o1}(p,q)$ phase.

By using the $\psi_{\rm in}(q)$ factor of (\ref{tDo}), where $q_0/K=0$, and transforming (\ref{Do12}) back to the direct space, we obtain (see the supplementary material)
\begin{equation}\begin{aligned}\label{moiré}
 &\psi_{o1}^{\rm out}(y) \propto \exp\left( -\frac{(y-y_D)^2}{l_D^2} +
 \frac{\rmi K_z y^2}{2r_D} + \frac{2\pi\rmi y}{\Lambda_\rho}  -\rmi h\rho\Delta y \right) , \\
 &\psi_{o2}^{\rm out}(y) \propto \exp\left( -\frac{y^2}{l_D^2} + \frac{\rmi K_z y^2}{2r_D} \right) ,
\end{aligned}\end{equation}
where $y_D=2\rho (t_A+z_{DA}) \tan(\Theta_B)$ is the offset between the interfering plane-wave components at the detector,
\begin{equation}\label{spacing-1}
 \Lambda_\rho = \frac{d}{\rho} \frac{K_z^2 z_D^2 \tan^4(\vartheta_0) + 4}{K_z^2 z_A z_D \tan^4(\vartheta_0) + 4} ,
\end{equation}
$d$ is the diffracting-plane spacing and $\vartheta_0$, $l_D$, and $1/r_D$ are the beam (vertical) divergence and the beam radius and wavefront curvature at the detection plane $z=z_D$ (see the appendix \ref{appendix:2}).

\subsubsection*{Fringe spacing.}
The superposition of the waves given in the equations (\ref{moiré}) yields interference fringes parallel to the $x$ axis and having $\Lambda_\rho$ spacing. When $z_D=z_A$ (that is, the detector is located at the analyser) or $\vartheta_0 \rightarrow 0$ (that is, the incoming wave is plane), the fringe spacing is
\begin{subequations}\begin{equation}
 \Lambda_\rho\big|_{\substack{z_D=z_A \\ \vartheta_0 \rightarrow 0}} = d \big/ \rho .
\end{equation}
Also, if $\vartheta_0 \rightarrow \pi\big/2$ (that is, the incoming wave is spherical), the fringe spacing is
\begin{equation}\label{lambda-rho}
 \Lambda_\rho\big|_{\vartheta_0 \rightarrow \pi/2} = \frac{d}{\rho} \frac{z_D}{z_A} .
\end{equation}\end{subequations}

\subsubsection*{Fringe contrast.}

By using the equations (\ref{moiré}), the contrast of the fringe pattern is (see the supplementary material)
\begin{equation}\label{contrast-1}
 \Gamma = \frac{2\big| \psi_{o1}(y) \psi_{o2}^*(y) \big|}{|\psi_{o1}(y)|^2+|\psi_{o2}(y)|^2} = \mbox{sech}\left[ \frac{y_D(2y-y_D)}{l_D^2} \right] ,
\end{equation}
where $l_D$ is the vertical radius of the beam at the detector. Figure \ref{fig04}, where $l_D$ must substitute for $w_D$, shows how the contrast depends on $y$ and $y_D$ (see also the section \ref{a:misalignment:incoherent}).

When $y_D/l_D=0$ (that is when the analyser is aligned to the splitter/mirror pair or the detection plane is at the exit surface of the analyser) and when $y=y_D/2$, we achieve the maximum contrast. If the analyser is misaligned and the detection plane is far from the analyser, that is, if $y_D/l_D \ne 0$, as $|y-y_D/2|$ increases the fringe contrast decays approximately like a Gaussian having $3l_D^2 \big/(5 y_D)$ standard deviation.

\subsubsection{Travelling fringes.}

The analyser displacement $u$ retards or advances by $hu$ the phase of the waves reflected by the analyser. As a result, a scanning analyser gives rise to travelling fringes, the period of which is equal to the spacing of the diffracting planes. This is the foundation of the measurement of the silicon lattice parameter by combined x-ray and optical interferometry \cite{Massa_2011}.

If the detectors do not resolve the interference pattern, but counts the total particle count, integrations are necessary to obtain the observed signals $I_n$. Hence,
\begin{equation}\label{signal}
 I_n = \int_{-\infty}^{+\infty} \!\! |\tpsi_{n1}(\bp) + \tpsi_{n2}(\bp)|^2 \,\rmd \bp
 = J_n \big[ 1 + \Gamma_n \cos(\Phi_n) \big] ,
\end{equation}
where we assumed an infinite detector aperture and carried out the integration in the reciprocal space.

In the x-ray case, since photons by conventional sources have any polarisation, with equal probability, we add the $\sigma$ and $\pi$ polarisations incoherently, which is unnecessary in the neutron case. Therefore, in (\ref{signal}),
\begin{subequations}\label{Xioh_coherent}\begin{eqnarray}
 J_n         &= &\sum_{\substack{ {\beta=\sigma,\pi} \\ {i=1,2} }} |A(t_i)|^2 \int_{-\infty}^{+\infty} |\tpsi_{n,i}^\beta(\bp)|^2\, \rmd \bp , \\
 \Xi_n       &= &A(t_1)A^*(t_2) \sum_{\beta=\sigma,\pi} \int_{-\infty}^{+\infty} \tpsi_{n,1}^\beta(\bp) \tpsi_{n,2}^{\beta *}(\bp)\, \rmd \bp , \; \\
 \Gamma_n    &= &2|\Xi_n|\big/J_n , \\
 \Phi_n      &= &\arg(\Xi_n) ,
\end{eqnarray}\end{subequations}
where the star indicates complex conjugation and the $hu$ phase yielded by the analyser displacement is included in $\Phi_n$. The first order systematic errors in the measurement of the silicon lattice parameter by scanning x-ray interferometry were investigated in \cite{Vittone:1997b}.

By using the $\psi_{\rm in}(q)$ factor of (\ref{tDo}), where $q_0/K=0$, considering the leaving $o$ state, ideal geometry (i.e., $t_A=t_S$, $t_{M1}=t_{M2}$, $\theta=0$ rad), neglecting the $q\rho$ offset of the analyser Bragg alignment, and carrying out the integrations over $q$, we obtain (see the supplementary material)
\begin{equation}\label{visibility}
 \Gamma = \frac{2|\Xi_o|}{J_o} \propto \exp\left( -\frac{y_S^2}{2l_0^2} - \frac{h^2l_0^2\rho^2}{8} \right)
        \exp\left( -\frac{h^2\rho^2 l_A^2}{8} \right) ,
\end{equation}
where $l_A$ is the (vertical) beam size at the analyser. In (\ref{visibility}), the $y_S^2/(2l_0^2)$ term originates from the different intensities of the interfering rays, whereas the $h^2l_0^2\rho^2/8$ term originates from the tilt of the interfering wavefronts.

\subsection{Interference signal: partially coherent source}\label{s:incoherent}

When the beam particles are in the mixed state $j_{\rm in}$, calculating the densities of the leaving particles,
\begin{subequations}\begin{eqnarray}\label{p-spectra}
 \tS_n(\bp) &= &\langle \bp, n | j_{\rm out} | n,\bp \rangle = \tj_{nn}(p,q) , \\ \nonumber
 S_n(\bx)   &= &\langle \bx, n | j_{\rm out} | n,\bx \rangle = j_{nn}(x,y)
 = \frac{1}{4\pi^2} \int_{-\infty}^{+\infty} \tj_{nn}(\bp,\bp') \rme^{+\rmi\bp\cdot\bx} \rme^{-\rmi\bp'\cdot\bx}\,
            \rmd\bp\, \rmd\bp'  \label{Snx} ,
\end{eqnarray}\end{subequations}
requires propagating $j_{\rm in}$ through the interferometer.

This propagation is given by $j_{\rm out} = X j_{\rm in} X^\dag$, where the dagger indicates the adjoint and $X$ is given by (\ref{X}) or, by using the reciprocal-space representation,
\begin{equation}\label{jout-1}
 \langle \bp_1 | j_{\rm out} | \bp_2 \rangle = \int_{-\infty}^{+\infty}
 \langle \bp_1 |X| \bp'\rangle\langle \bp' |j_{\rm in}| \bp'' \rangle\langle \bp'' |X^\dag| \bp_2 \rangle\, \rmd\bp' \rmd\bp''
 = \int_{-\infty}^{+\infty} \langle \bp_1 |XP_o| \bp'\rangle \tj_{\rm in}(\bp', \bp'') \langle \bp'' |(XP_o)^\dag| \bp_2 \rangle \, \rmd\bp' \rmd\bp'' ,
\end{equation}
where $\langle \bp' |j_{\rm in}| \bp'' \rangle$ is given by (\ref{jin2}), we took $P_o^2=P_o$ and $P_o^\dag = P_o$ into account, and the scalar-valued function $\tj_{\rm in}(\bp', \bp'')$ is given by (\ref{Jp12}). The representations of $XP_o$ and $(XP_o)^\dag$ are
\begin{subequations}\begin{equation}
 \langle \bp |XP_o| \bp'\rangle = \left( \begin{array}{cc}
                                      \tX_o(\bp,\bp') & 0 \\
                                      \tX_h(\bp,\bp') & 0 \\
                                    \end{array} \right) ,
\end{equation}
and
\begin{equation}
 \langle \bp'' |(XP_o)^\dag| \bp\rangle =
 \left( \begin{array}{cc}
 \tX_o^*(\bp,\bp'') &  \tX_h^*(\bp,\bp'') \\
 0                  & 0 \\
 \end{array} \right) ,
\end{equation}\end{subequations}
where, omitting unessential shared phases,
\begin{equation}\label{Xni}
 \tX_n(\bp,\bp') = \exp\left( -\frac{\rmi(p^2 + q^2)z_D}{2K_z} \right) \sum_{i=1,2} A(t_i)\tX_{ni}(\bp,\bp') ,
\end{equation}
and $\tX_{ni}(\bp,\bp')$ are given in the appendix \ref{appendix:4}. The $nm$ elements of (\ref{jout-1}) are
\begin{equation}\label{jout-3}
 \tj_{nm}(\bp_1,\bp_2) = \int_{-\infty}^{+\infty} \tX_n(\bp_1,\bp') \tX_m^*(\bp_2,\bp'') \tj_{\rm in}(\bp', \bp'')\, \rmd\bp' \rmd\bp''
 = \exp\left( -\frac{\rmi(|\bp_1|^2 - |\bp_2|^2)z_D}{2K_z} \right) \sum_{\substack{ {i=1,2} \\ {j=1,2} }} A(t_i)A^*(t_j) \tj_{nm}^{(ij)}(\bp_1,\bp_2) .
\end{equation}

By introducing $p_- = p_1 - p_2$ and $q_- = q_1 - q_2$, the diagonal terms $\tj_{nn}^{(ij)}(\bp_1,\bp_2)$ needed to calculate the particle densities are (see the supplementary material)
\begin{subequations}\label{tjoo12}\begin{eqnarray}
    \tj_{oo}^{(11)}(\bp_1,\bp_2)  &= &T(p_1;t_S)R(p_1;t_{M1})R(p_1+\theta K_z-\rho q_1;t_A)T^*(p_2;t_S)R^*(p_2;t_{M1})R^*(p_2+\theta K_z-\rho q_2;t_A) \\ \nonumber
 & &\times \tj_{\rm in}(p_1,q_1-h\rho,p_2,q_2-h\rho) \exp\left[ -\rmi ( p_-x_1 - q_-y_S ) \right] ,\\
    \tj_{oo}^{(22)}(\bp_1,\bp_2)  &= &R(p_1;t_S)R(p_1;t_{M2})T(p_1+\theta K_z-\rho q_1;t_A)R^*(p_2;t_S)R^*(p_2;t_{M2})T^*(p_2+\theta K_z-\rho q_2;t_A) \\ \nonumber
 & &\times \tj_{\rm in}(p_1,q_1,p_2,q_2) \exp( -\rmi p_-x_2 ) ,\\
    \tj_{oo}^{(12)}(\bp_1,\bp_2) &= &T(p_1;t_S)R(p_1;t_{M1})R(p_1+\theta K_z-\rho q_1;t_A)R^*(p_2;t_S)R^*(p_2;t_{M2})T^*(p_2+\theta K_z-\rho q_2;t_A) \\ \nonumber
 & &\times \tj_{\rm in}(p_1,q_1-h\rho,p_2,q_2) \exp\left\{ -\rmi \left[ p_1x_1 - p_2x_2 - q_1y_S + h(u+\theta\Delta z) \right]\right\} ,\\
    \tj_{oo}^{(21)}(\bp_1,\bp_2) &= &\tj_{oo}^{(12)*}(\bp_2,\bp_1) ,\\
    \tj_{hh}^{(11)}(\bp_1,\bp_2)  &= &T(p_1;t_S)R(p_1;t_{M1})T(-p_1-\theta K_z+\rho q_1;t_A)T^*(p_2;t_S)R^*(p_2;t_{M1})T^*(-p_2-\theta K_z+\rho q_2;t_A) \\ \nonumber
 & &\times \tj_{\rm in}(p_1,q_1,p_2,q_2) \exp( -\rmi p_-x_1 ) ,\\
    \tj_{hh}^{(22)}(\bp_1,\bp_2)  &= &R(p_1;t_S)R(p_1;t_{M2})R(p_1+\theta K_z-\rho q_1;t_A)R^*(p_2;t_S)R^*(p_2;t_{M2})R^*(p_2+\theta K_z-\rho q_2;t_A) \\ \nonumber
 & &\times \tj_{\rm in}(p_1,q_1+h\rho,p_2,q_2+h\rho) \exp\left[ -\rmi ( p_-x_2 + q_-y_S ) \right] ,\\
    \tj_{hh}^{(12)}(\bp_1,\bp_2) &= &T(p_1;t_S)R(p_1;t_{M1})T(-p_1-\theta K_z+\rho q_1;t_A)R^*(p_2;t_S)R^*(p_2;t_{M2})R^*(p_2+\theta K_z-\rho q_2;t_A) \\ \nonumber
 & &\times \tj_{\rm in}(p_1,q_1,p_2,q_2+h\rho) \exp\left\{ -\rmi \left[ p_1x_1 - p_2x_2 - q_2y_S + h(u+\theta\Delta z) \right]\right\} ,\\
    \tj_{hh}^{(21)}(\bp_1,\bp_2) &= &\tj_{hh}^{(12)*}(\bp_2,\bp_1) .
\end{eqnarray}\end{subequations}
We remember that $t_{1,2}$ are total crystal thickness along the 1 and 2 arms, $z_D$ is the detector distance from the source, $x_{1,2}$ are the horizontal distances from the origin of the beams leaving the interferometer, and $y_S$ is the start offset between the wave packets interfering collinearly, see Fig.\ \ref{fig01}.

\subsubsection{Triple Laue rocking curve.}

The triple Laue rocking curve is given by
\begin{subequations}\begin{equation}
 I_{\rm RRR}(\theta,\rho) \propto \int_{-\infty}^{+\infty} \big| \tj_{hh}^{(22)}(p,q) \big|^2\, \rmd p\, \rmd q
 \propto \int_{-\infty}^{+\infty} \big| R(p+\theta K_z-\rho q;t_A)  R(p;t_{M2}) R(p;t_S) \big|^2
         \tj_{\rm in}(p,q+h\rho) \, \rmd p\, \rmd q ,
\end{equation}
where, see (\ref{Jp12}),
\begin{equation}\label{tjinpq}
 \tj_{\rm in}(p,q) \propto \exp\left[ -\frac{(p^2+q^2)\ell_0^2}{2} \right] .
\end{equation}\end{subequations}

Since $\tj_{\rm in}(p,q)$ equals $|\tpsi_{\rm in}(p,q)|^2$ (see (\ref{tDo}), where $p_0$ and $q_0$ are set to zero), the rocking curves shown in Fig.\ \ref{fig03} hold true also for an incoherent neutron source. However, the visibility of the pendell\"osung fringes is not ruled by the source size $w_0$, but by the coherence length $\ell_0$; the smaller, the lesser the fringe visibility.

\subsubsection{Moiré fringes.}\label{a:misalignment:incoherent}

The moiré pattern yielded by the analyser rotation $\rho$ is encoded by the particle density (\ref{Snx}). Let us neglect the $q\rho$ offset of the analyser
Bragg alignment. Considering only the $o$ state and $q$-factor, the addenda of $\tj_{oo}(q_1,q_2)$, see (\ref{jout-3}) and (\ref{tjoo12}), are
\begin{subequations}\label{jnmq}\begin{eqnarray}
 \tj_{oo}^{(11)}(q_1,q_2)\hspace{-2mm}   &\propto\hspace{-2mm} &\tj_{\rm in}(q_1-h\rho,q_2-h\rho) \rme^{\rmi q_- y_S - \frac{\rmi (q_1^2-q_2^2)z_D}{2K_z}} ,\; \\
 \tj_{oo}^{(22)}(q_1,q_2)\hspace{-2mm}   &\propto\hspace{-2mm} &\tj_{\rm in}(q_1,q_2) \rme^{-\frac{\rmi (q_1^2-q_2^2)z_D}{2K_z}} ,\\
 \tj_{oo}^{(12)}(q_1,q_2)\hspace{-2mm}   &\propto\hspace{-2mm} &\tj_{\rm in}(q_1-h\rho,q_2) \rme^{\rmi q_1 y_S - \frac{\rmi (q_1^2-q_2^2)z_D}{2K_z}} , \\
 \tj_{oo}^{(21)}(q_1,q_2)\hspace{-2mm}   &=\hspace{-2mm} &\tj_{oo}^{(12)^*}(q_2,q_1) ,
\end{eqnarray}\end{subequations}
where $\tj_{\rm in}(q_1,q_2)$ is the $q$ factor of (\ref{Jp12}), $y_S=2\rho z_A \tan(\Theta_B)$ is the start offset between the particles interfering collinearly (see Fig.\ \ref{fig01}), and we included the $(q_1^2-q_2^2)z/(2K_z)$ phase picked from (\ref{jout-3}).

The results of the integrations (\ref{Snx}) are (see the supplementary material)
\begin{subequations}\label{jnmy}\begin{eqnarray}
    j_{oo}^{(11)}(y)\hspace{-2mm}    &\propto\hspace{-2mm} &\exp\left[-\frac{2(y-y_D)^2}{w_D^2} \right] ,\\
    j_{oo}^{(22)}(y)\hspace{-2mm}    &\propto\hspace{-2mm} &\exp\left(-\frac{2y^2}{w_D^2} \right) ,\\
    j_{oo}^{(12)}(y)\hspace{-2mm} &\propto\hspace{-2mm} &\exp\left[
    -\frac{y_D^2}{2\ell_D^2} - \frac{y^2+(y-y_D)^2}{w_D^2} + \frac{2\pi\rmi y}{\Lambda_\rho} \right] ,\hspace{2mm} \\
    j_{oo}^{(21)}(y)\hspace{-2mm} &=\hspace{-2mm} &j_{oo}^{(12)*}(y) .
\end{eqnarray}\end{subequations}
where $\ell_D$ and $w_D$ are the correlation length and the beam size at the detection plane, respectively,
\begin{equation}\label{spacing-2}
 \Lambda_\rho = \frac{d}{\rho}
 \frac{K_z^2 z_D^2 \tan^2(\vartheta_\ell)\tan^2(\vartheta_w) + 4}{K_z^2 z_A z_D \tan^2(\vartheta_\ell)\tan^2(\vartheta_w) + 4} ,
\end{equation}
is the spacing of the moiré fringes, $\tan(\vartheta_\ell) = 2/(K_z\ell_0)$, and $\tan(\vartheta_w) = 2/(K_zw_0)$.

\begin{figure}\centering
\includegraphics[width=0.5\columnwidth]{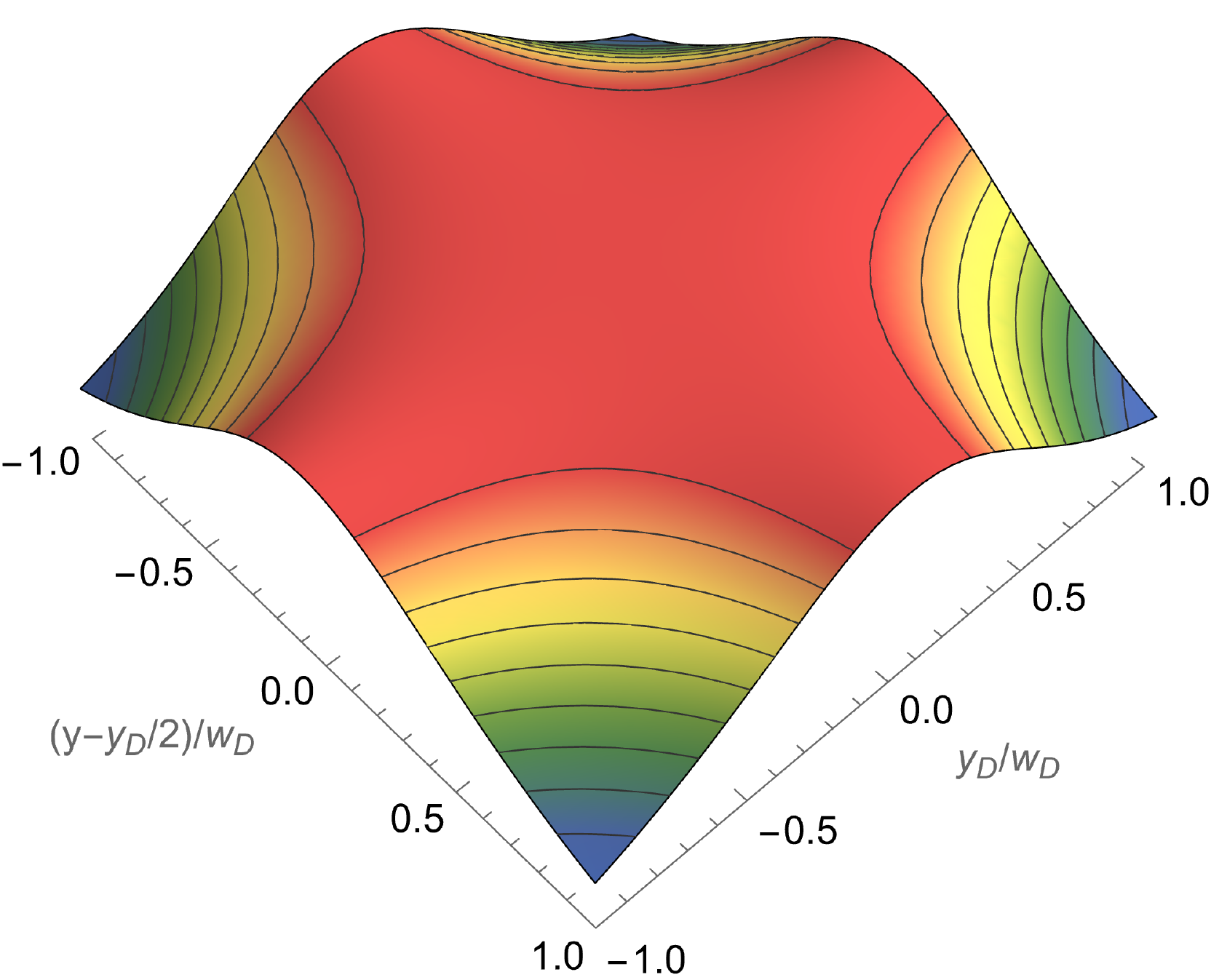}
\caption{ Normalised contrast $\Gamma$, see (21), or $\exp\big[y_D^2\big/(2\ell_D^2)\big] \Gamma$, see (36), of the moiré fringes due to the pitch angle of the analyser calculated for a Gaussian wavefunction [see (21)] or a Gaussian Schell-model of the source [see (36)]. The color scale is from red, 1, to blue, 0.27. $\ell_D$, $w_D$, and $y_D$ are the coherence length, radius, and offset of the interfering beams at the detector, respectively (see Fig.\ 1).} \label{fig04}
\end{figure}

\subsubsection*{Fringe contrast.}

Putting (\ref{Snx}) and the equations (\ref{jnmy}) together, the $y$-factor of the ($o$ state) particle density is
\begin{equation}
 S_o(y) \approx 2\rme^{-2y^2\big/w_D^2} \big[ 1 + \Gamma \cos(2\pi y\big/\Lambda_\rho) \big] ,
\end{equation}
where
\begin{equation}\label{contrast}
 \Gamma = \frac{2|j_{oo}^{(12)}|}{j_{oo}^{(11)}+j_{oo}^{(22)}} = \exp\left(-\frac{y_D^2}{2\ell_D^2}\right) {\rm{sech}}\left( \frac{y_D(2y-y_D)}{w_D^2} \right) ,
\end{equation}
which, apart from the $\exp\big[-y_D^2\big/(2\ell_D^2)\big]$ factor, is the same as (\ref{contrast-1}) and tends to it as $\ell_0/w_0 \rightarrow\infty$.

The contrast of the fringe pattern is shown in Fig.\ \ref{fig04}. The maximum, which is equal to
\begin{equation}\label{max:contrast}
 \Gamma_{\rm max} = \exp\big[ -y_D^2 \big/ (2\ell_D^2) \big] ,
\end{equation}
occurs when $y = y_D\big/2$, that is, when $y$ is midway between the interfering-beam spots, $w_D/y_D \rightarrow\infty$, and $y_D = 0$, in which case (and only in this last case) it is one.

According to (\ref{max:contrast}), the maximum contrast is set by the coherence length $\ell_D$ at the detection plane. Because of the van Cittert-Zernike theorem, the coherence length increases with the detector distance from the source. For instance, when $\ell_0/w_0 \rightarrow 0$ (that is, the source is completely incoherent), the coherence length at the detection plane is $\ell_D = 4z\big/(K_z w_0)$, see (\ref{lcz}). Therefore, reducing the source radius $w_0$ increases the contrast, as it is  reported in \cite{Lemmel_2021,S18data}.

We can qualitatively understand the effect of the source coherence as follows. According to (\ref{Do12}) and the therein discussion, the phase difference of fringe patterns originated by any pair of point-like sources vertically spaced by $\Delta y$ is $h\rho\Delta y$. If $z_D=z_A$ (that is, the detection plane is the analyser), the patterns' spacing is $\Delta y$ and their phases run as $h\rho y$. Therefore, the fringes overlap in phase and their contrast is one, as predicted by (\ref{contrast}). Contrary, if $z_D \ne z_A$, according to (\ref{lambda-rho}), the patterns' spacing increases to $\Delta y+y_D$ and their phases run as $h\rho z_A y/z_D$. Therefore, the fringes overlap out of phase and we lose contrast.

\subsubsection{Travelling fringes.}

The integration of the particle densities (\ref{p-spectra}) over $\bp$,
\begin{equation}\label{s-incoherent}
 I_n = \int_{-\infty}^{+\infty} \tS_n(\bp)\, \rmd \bp = J_n \big[ 1 + \Gamma_n \cos(\Phi_n) \big] ,
\end{equation}
gives the counts of total particle. By noticing that $\tj_{nn}^{(ii)}(\bp,\bp)$ are reals and $\tj_{nn}^{(21)}(\bp,\bp)=\tj_{nn}^{(12)*}(\bp,\bp)$, we obtain (the sums over the $\sigma$ and $\pi$ polarisations are irrelevant in the neutron case)
\begin{subequations}\label{Xioh_incoherent}\begin{eqnarray}
 J_n         &= &\sum_{\substack{ {\beta=\sigma,\pi} \\ {i=1,2} }} \int_{-\infty}^{+\infty} \tj_{nn}^{(ii)\beta}(\bp,\bp)\, \rmd \bp , \\
 \Xi_n       &= &\sum_{\beta=\sigma,\pi} \int_{-\infty}^{+\infty} \tj_{nn}^{(12)\beta}(\bp,\bp)\, \rmd \bp , \label{Xioh_incoherent_b}
\end{eqnarray}\end{subequations}
$\Gamma_n=2|\Xi_n|/J_n$, and $\Phi_n=\arg(\Xi_n)$. The expressions of $\tj_{nn}^{(i)}(\bp,\bp)$ and $\tj_{nn}^{(ij)}(\bp,\bp)$ are given in the appendix \ref{appendix:5}. It is worth noting that they are the equivalents of $|\tpsi_{n,i}(\bp)|^2$ and $ \tpsi_{n,1}^\beta(\bp) \tpsi_{n,2}^*(\bp)$, respectively, in the equations (\ref{Xioh_coherent}).

The joint effect of the analyser pitch angle and incoherent source on the fringe visibility can be investigated observing that the density matrix $j_{\rm in}$ is separable, see the section  (\ref{incoherent}). Considering the leaving $o$ state, ideal geometry (i.e., $t_A=t_S$, $t_{M1}=t_{M2}$, $\theta=0$ rad), neglecting the $q\rho$ offset of the analyser Bragg alignment, and carrying out the integrations over $q$, we obtain (see the supplementary material)
\begin{equation}
 \Gamma = \exp\left( -\frac{y_S^2}{2\ell_0^2} -\frac{h^2\rho^2 w_0^2}{8} \right) = \exp\left( -\frac{h^2 w_A^2 \rho^2}{8} \right) ,
\end{equation}
where $w_A$ is the (vertical) size of the particle density at the analyser location, see (\ref{wcz}).

This visibility is the same as (\ref{visibility}). In the same way as in (\ref{visibility}), the $y_S^2/(2\ell_0^2)$ term originates from the different intensities of the rays interfering coherently, whereas the $h^2w_0^2\rho^2/8$ term originates from the tilt of the interfering wavefronts. If $h\ell_0 \rightarrow 0$, then $hw_A \rightarrow \infty$ and $\Gamma \rightarrow 0$. Consequently, if the particles' source is completely incoherent, the interferometer operates only with a perfectly aligned analyser.

\section{Conclusions}

The proof-of-principle demonstration that the alignment and operation of a split-crystal interferometer with the accuracy required for neutron interference are technically possible \cite{Lemmel_2021,Massa:2022} prompts the design of split-crystal skew-symmetric interferometers operating with both x-rays and neutrons and having the potential of crystal separations up to the meter scale.

Quantifying the effect of the misalignments between the crystals on the visibility and phase of the interference signals is essential to identify the specifications necessary to successful manufacture and operate the interferometer. Three dimensional operation and spatial coherence play an important role in determining the interference visibility. Therefore, otherwise from previous studies, the paper novelty is a formalism to model split-crystal interferometers in three dimensions and operating both with coherent and partially coherent x-rays and neutrons.

For the sake of algebraic simplicity, we considered a symmetric geometry and only the analyser free to move with respect to the splitter-mirror pair. A split skew-symmetric geometry, where one of the two mirrors is integral with the analyser and free to move with respect to the other, which is integral with the splitter, can be studied along the same lines and will be the subject matter of future investigations.

We quantified the dependence of the pendell\"osung-fringe phase on the out-of-reflection-plane propagation of the plane-wave components of the particles travelling through the interferometer. If the analyser's pitch angle is misaligned, the averaging of the out-of-phase pendell\"osung fringes associated with different plane-wave components reduces the fringe visibility, which reduction can be used to approach the right alignment. Eventually, the visibility of the pendell\"osung and moiré fringes as a function of the analyser pitch angle delivers information about the source coherence.

Our formalism will also allow the effects of parasitic pitch rotations -- associated with the analyser axial displacement -- on the phase of the travelling fringes to be quantified. This quantification, which is integral to the investigation of systematic effects in the measurement of the $^{28}$Si lattice parameter, will be the subject matter of a future investigation.

Eventually, the developed formalism opens new possibilities to study gravitationally-induced interference in a neutron interferometer from first principles \cite{Bonse:1984,Horne:1986,Werner:1988}.

\appendix
\numberwithin{equation}{section}

\section{Reciprocal space representation}\label{appendix:1}

According to the sign convention adopted in the propagation of electromagnetic waves, the direct space representation of the state $|\psi\rangle$ is the plane-wave superposition
\begin{equation*}
 \langle \bx, n |\psi(z)\rangle = \psi_n(\bx;z) = \frac{1}{2\pi} \int_{-\infty}^{+\infty} \tpsi_n(\bp;z)\rme^{\rmi \bp\cdot\bx}\, \rmd \bp ,
\end{equation*}
where
\begin{equation*}
 \langle \bp, n |\psi(z)\rangle = \tpsi_n(\bp;z) = \frac{1}{2\pi} \int_{-\infty}^{+\infty} \psi_n(\bx;z)\rme^{-\rmi \bp\cdot\bx}\, \rmd \bx
\end{equation*}
is the reciprocal space representation. We also use the orthogonality and completeness (in the Dirac sense) of the
\begin{equation*}
 \langle \bx|\bp \rangle = \rme^{+\rmi\bp\cdot\bx} \big/ (2\pi)
\end{equation*}
and
\begin{equation*}
 \langle \bp|\bx \rangle = \rme^{-\rmi\bp\cdot\bx} \big/ (2\pi)
\end{equation*}
bases, which are expressed by the integral representations of the delta distribution
\begin{equation*}
 \langle \bp|\bp' \rangle = \delta(\bp'-\bp) = \frac{1}{4\pi^2} \int_{-\infty}^{+\infty} \rme^{\rmi(\bp'-\bp)\cdot\bx}\, \rmd \bx
\end{equation*}
and
\begin{equation*}
 \langle \bx|\bx' \rangle = \delta(\bx-\bx') = \frac{1}{4\pi^2} \int_{-\infty}^{+\infty} \rme^{\rmi\bp\cdot(\bx-\bx')}\, \rmd \bp .
\end{equation*}

\section{Laue diffraction in a displaced crystal}\label{appendix:3}

Since the analyser might slip and orient differently from the splitter-mirror block, we need representing the (quantum) state of the incoming particles in a roto-translated basis.

Any misalignment is described by a rotation about an arbitrary point plus a translation. Vertical displacements (along the $y$ axis) are irrelevant because the analyser is symmetrically cut and plane-parallel. Displacements along the $z$ axis are encoded by the crystal spacings. Eventually, it is timely that, in the rotation centre, the $o$ and $h$ bases have the same phase. Consequently, the interferometer focus, $\rb_F=(x_F,0,z_F)$, is the optimal choice and the displacement $u$ along the $x$ axis is the only additional degree of freedom (see Fig.\ \ref{fig01}). Since the interferometer is insensitive to the roll angle, which plays a role only if it is macroscopic, we neglect it.

After translating the interferometer and analyser origins in $\rb_F$, the interferometer position-vector $\rb_L$ (relative to $\rb_F$) is seen from the analyser as
\begin{equation*}
 \rb_A = M\rb_L = R(\rb_L) + u\hx_L ,
\end{equation*}
where
\begin{equation*}
 R = \left( \begin{array}{ccc}
            1       &-\rho    &\theta \\
            \rho    &1        &0 \\
            -\theta &0        &1
            \end{array} \right) ,
\end{equation*}
and $\theta$ and $\rho$ indicate yaw (the Bragg's rotation about the vertical, $y$, axis) and pitch (the rotation about the $z$ axis) angles, respectively. Accordingly, the analyser, first, rotates about $\rb_F$ and, then, translates by $-u\hx_L$. From the analyser viewpoint, the splitter-mirror block counter-rotates about $\rb_F$ and translates in the $u\hx_A$ direction.

We indicates by $\hM$ the linear operator associated to $M$ that changes the abstract single-particle state leaving the first crystal, $|\psi^L(z_L)\rangle$, to that seen by the analyser, $|\psi^A(z_A) \rangle$, that is, $|\psi^A(z_A) \rangle=\hM|\psi^L(z_L) \rangle$. Since the wave-function value at any given point is unchanged despite the change of the reference frame, $ \rb_A = M\rb_L$, the operator $\hM$ that we are seeking can be found by explicit construction. Therefore, by using (\ref{Doh}),
\begin{equation*}
 | o \rangle_L = \rme^{\rmi \bK_o\cdot\rb_L} \left( \begin{array}{c} 1 \\ 0 \end{array} \right)_L =
 \rme^{\rmi \bK_o\cdot(M^{-1}\rb_A)} \left( \begin{array}{c} 1 \\ 0 \end{array} \right)_L
\end{equation*}
and, since the analyser does not see $\bK_n$ rotations,
\begin{equation*}
 | o \rangle_A = \rme^{\rmi \bK_o\cdot\rb_A} \left( \begin{array}{c} 1 \\ 0 \end{array} \right)_A =
 \rme^{\rmi \bK_o\cdot\rb_A} \hM \left( \begin{array}{c} 1 \\ 0 \end{array} \right)_L
\end{equation*}
Since$|n\rangle_L = |n\rangle_A$, the $\hM$ restriction to $V_2$ is
\begin{equation*}
 _L\langle m |\hM | n \rangle_L = \exp[\rmi \bK_n\cdot(M^{-1}\rb_A-\rb_A)] \delta_{mn} ,
\end{equation*}
where, by using (\ref{Koh}),
\begin{equation}\label{C1}
 \bK_{o,h}\cdot(M^{-1}\rb_A-\rb_A) = \theta K_z x'_A \pm h(u -\rho y_A +\theta z_A)\big/2 ,
\end{equation}
$K_z = K\cos(\Theta_B)$, $K_x = K\sin(\Theta_B)=h\big/2$, and $x_A-s$ has been redefined as $x'_A$.

Similarly, since $\langle \rb_L |\psi_n^L \rangle = \langle \rb_A | \psi_n^A \rangle = \langle \rb_A |\hM|\psi_n^L\rangle$, to find the $\hM$ restriction to $L_2$, we observe that
\begin{equation*}
 \langle \rb_A | \psi_n^A \rangle = \langle \rb_A | \hM |\psi_n^L \rangle =  \langle M^{-1}\rb_A |\psi_n^L \rangle = \psi_n^L(M^{-1}\rb_A) .
\end{equation*}
Consequently,
\begin{equation*}
 \langle \rb_A | \hM |\psi_n^L \rangle = \int_{-\infty}^{+\infty}
 \langle \rb_A | \hM | \rb_L \rangle\langle \rb_L |\psi_n^L \rangle \, \rmd \rb_L 
 = \int_{-\infty}^{+\infty}\langle \rb_A | \hM | \rb_L \rangle \psi_n^L(\rb_L) \, \rmd \rb_L = \psi_n^L(M^{-1}\rb_A)
\end{equation*}
and
\begin{equation}\label{C2}
 \langle \rb_A | \hM | \rb_L \rangle = \delta(M^{-1}\rb_A-\rb_L) 
 \approx \delta( x'_A -x_L -\theta z_A +\rho y_A, y_A -y_L -\rho x'_A, z_A -z_L +\theta x'_A ) ,
\end{equation}

Owing to small horizontal extensions of $\psi_{o,h}^A(\rb_A)$ about $x_{2,1}^A-x_F^A$, we can approximate $x_A$ as $x_2^A-x_F^A$ ($o$ beam) and $x_1^A-x_F^A$ ($h$ beam), see Fig.\ \ref{fig01}. Therefore, since the interferometer operation requires $(x_{2,1}^A-x_F^A)/\Delta_e \propto \Delta z/\Delta_e \ll 1$ \cite{Vittone:1997b}, $\theta x'_A$ is a second-order term and will be neglected from now on. Hence,
\begin{equation*}
 \langle \rb_A | \hM | \rb_L \rangle =  \hM(\bx_A,\bx_L;z_A)\delta(z_A -z_L) .
\end{equation*}

Putting it all together and setting $z=z_A=z_L$ (see the supplementary material), the direct- and reciprocal-space representations of $\hM$ are
\begin{eqnarray}\label{M}
 \hM(\bx_A,\bx_L;z)\hspace{-2mm} &=\hspace{-2mm} &\rme^{\rmi K_z\theta x'_A} \left( \begin{array}{cc}
            \rme^{\rmi h(u-\rho y_A+\theta z)/2} &0 \\
            0 &\rme^{-\rmi h(u-\rho y_A+\theta z)/2}
                                          \end{array} \right) \delta( x'_A -x_L +\rho y_A -\theta z, y_A -y_L -\rho x'_A) , \\ \nonumber
 \hM(\bp',\bp;z)\hspace{-2mm} &=\hspace{-2mm} &\int_{-\infty}^{+\infty}
 \langle \bp'|\bx' \rangle \langle z,\bx' |\hM | \bx,z \rangle \langle \bx | \bp \rangle \, \rmd \bx' \rmd \bx  \\
  &=\hspace{-2mm} &\left( \begin{array}{cc}
            \rme^{+\rmi h(u+z\theta)/2}\delta( q-q'-h\rho/2) & 0 \\
            0 &\rme^{-\rmi h(u+z\theta)/2}\delta( q-q'+h\rho/2)
            \end{array} \right) \delta(p-p'-\rho q'+\theta K_z) ,
\end{eqnarray}
\noindent where $\bp'=(p',q')$ and $\bp=(p,q)$ are the variables conjugate to $(x'_A, y_A)$ and $(x_L-x_F, y_L-y_F)$, respectively, we neglected the second-order terms proportional to $p\theta$ and $p\rho$, and approximated $\rho q = \rho q' \pm h\rho^2\big/2$ by $\rho q'$. The analyser representation of $|\psi_L(z)\rangle$ is (see the supplementary material)
\begin{eqnarray*}
 \tpsi_A(\bp';z) &= &\langle \bp' | \psi_A(z)\rangle = \langle \bp' |\hM(z) | \psi_L(z)\rangle =
 \int_{-\infty}^{+\infty} \langle \bp'|\hM(z)| \bp \rangle \langle \bp |\psi_L(z)\rangle \,\rmd \bp \\
 &= &\int_{-\infty}^{+\infty} \hM(\bp',\bp;z) \tpsi_L(\bp;z)\rangle \,\rmd \bp
 = \left( \begin{array}{c}  \tpsi_{Lo}(p'+\rho q'-\theta K_z, q'+h\rho\big/2;z) \rme^{+\rmi h(u+\theta z)/2} \\
                            \tpsi_{Lh}(p'+\rho q'-\theta K_z, q'-h\rho\big/2;z) \rme^{-\rmi h(u+\theta z)/2} \end{array} \right) .
\end{eqnarray*}
\noindent After propagating the state $| \psi_A(\Delta z)\rangle = \hM(\Delta z) | \psi_L(\Delta z) \rangle$ (where $|\psi_L(\Delta z) \rangle$ is the particle state at the $z = z_F+\Delta z$ plane, see Fig.\ \ref{fig01}) by the scattering matrix $U_0(t_A)$, the laboratory representation of $| \psi_A(\Delta z+t_A)\rangle = U_0(t_A) | \psi_A(\Delta z)\rangle$ is (see the supplementary material)
\begin{eqnarray*}
 \tpsi_L(\bp;\Delta z+t_A) = \langle \bp |\hM^{-1}(\Delta z+t_A) U_0(t_A) | \psi_A(\Delta z)\rangle =
 \int_{-\infty}^{+\infty} \hM^{-1}(\bp,\bp';\Delta z+t_A) U_0(p';t_A) \tpsi_A(\bp';\Delta z)\rangle \,\rmd \bp' \hspace{5mm}\\
 = \left( \begin{array}{c}
 \left[
 T(p-\rho q+\theta K_z;t_A)\tpsi_{Lo}(p, q;\Delta z) +R(p-\rho q+\theta K_z;t_A)\tpsi_{Lh}(p, q-h\rho;\Delta z) \rme^{-\rmi h(u+\theta\Delta z)}
 \right] \rme^{-\rmi(K_z\theta -q\rho)t_A\tan(\Theta_B)} \\[2mm]
 \left[
 R(p-\rho q+\theta K_z;t_A)\tpsi_{Lo}(p, q+h\rho;\Delta z) \rme^{+\rmi h(u+\theta\Delta z)} +T(-p+\rho q-\theta K_z;t_A)\tpsi_{Lh}(p, q;\Delta z)
 \right] \rme^{+\rmi(K_z\theta -q\rho)t_A\tan(\Theta_B)} \\
 \end{array} \right) \\
 \times\exp\left[ - \frac{\rmi (p^2+q^2) t_A}{2K_z} \right] ,
\end{eqnarray*}
\noindent where $\hM^{-1}(\Delta z+t_A)$ is obtained by the substitutions $\rho\rightarrow -\rho$, $\theta\rightarrow -\theta$, and $u \rightarrow -u$.

\section{Gaussian beam propagation}\label{appendix:2}

Let us consider a separable Gaussian beam that propagates in the $x-z$ plane at the $\Theta_B$ angle with respect to the $z$ axis. Its divergence $\theta_0$, radius $l_z$ and wavefront radius of curvature $r_z$ at the distance $z$ from the source are (see the supplementary material)
\begin{eqnarray*}
 \tan(\theta_0) &= &2/(K_z l_0) , \\
 l_z^2          &= &l_0^2 \left( 1 + \frac{4z^2}{K_z^2 l_0^4} \right), \\
 r_z            &= &\frac{K_z^2l_0^4 + 4z^2}{4z} = \frac{K_z^2l_0^2l_z^2}{4z} , \\
\end{eqnarray*}
where $l_0$ is the source radius measured in the $x-y$ plane and we omitted the indexes labelling the beam's $x$ and $y$ axes.

\section{Propagation of the density matrix} \label{appendix:4}

The matrix elements necessary to propagate the density matrix through the interferometer are given below (see the supplementary material). Unessential phases shared by the interfering $o1, o2$ and $h1, h2$ elements and second order terms have been omitted.
\begin{eqnarray*}
    \tX_{o1}(\bp,\bp')\hspace{-3mm} &=\hspace{-3mm} &R(p+\theta K_z-\rho q;t_A)  R(p;t_{M1}) T(p;t_S) \delta(p'-p,q'-q+h\rho)
 \exp\left\{ -\rmi \left[ px_1 - qy_S + h(u+\theta\Delta z) \right]\right\} ,\\
    \tX_{o2}(\bp,\bp')\hspace{-3mm} &=\hspace{-3mm} &T(p+\theta K_z-\rho q;t_A)R(p;t_{M2})R(p;t_S) \delta(p'-p,q'-q) \exp( -\rmi p x_2 ) ,\\
    \tX_{h1}(\bp,\bp')\hspace{-3mm} &=\hspace{-3mm} &T(-p-\theta K_z+\rho q;t_A)R(p;t_{M1})T(p;t_S) \delta(p'-p,q'-q) \exp( -\rmi p x_1 ) ,\\
    \tX_{h2}(\bp,\bp')\hspace{-3mm} &=\hspace{-3mm} &R(p+\theta K_z-\rho q;t_A)R(p;t_{M2})R(p;t_S) \delta(p'-p,q'-q-h\rho)
 \exp\left\{ -\rmi \left[ px_2 + qy_S - h(u+\theta\Delta z) \right]\right\} .
\end{eqnarray*}

\section{Particle densities} \label{appendix:5}

The diagonal elements of the propagated density matrix necessary to calculate the particle densities of the $o$ and $h$ states leaving the interferometer are (see the supplementary material)
\begin{eqnarray*}
    \tj_{oo}^{(11)}(\bp,\bp)  &= &\big|T(p_1;t_S)R(p;t_{M1})R(p+\theta K_z-\rho q;t_A)  \big|^2 \tj_{\rm in}(p,q-h\rho,p,q-h\rho) \\
    \tj_{oo}^{(22)}(\bp,\bp)  &= &\big| R(p;t_S)R(p;t_{M2})T(p+\theta K_z-\rho q;t_A) \big|^2 \tj_{\rm in}(p,q,p,q) \\
    \tj_{oo}^{(12)}(\bp,\bp)  &= &T(p;t_S)R(p;t_{M1})R(p+\theta K_z-\rho q;t_A) R^*(p;t_S)R^*(p;t_{M2})T^*(p+\theta K_z-\rho q;t_A) \\
 & &\times \tj_{\rm in}(p,q-h\rho,p,q) \exp\big[ -\rmi [px_0 - qy_S + h(u+\theta\Delta z)] \big] \\
    \tj_{oo}^{(21)}(\bp,\bp)  &= &\tj_{oo}^{(12)*}(\bp,\bp) \\
    \tj_{hh}^{(11)}(\bp,\bp)  &= &T(p;t_S)R(p;t_{M1})T(-p-\theta K_z+\rho q;t_A) \big|^2 \tj_{\rm in}(p,q,p,q) \\
    \tj_{hh}^{(22)}(\bp,\bp)  &= &\big| R(p;t_S)R(p;t_{M2})R(p+\theta K_z-\rho q;t_A) \big|^2 \tj_{\rm in}(p,q+h\rho,p,q+h\rho) \\
    \tj_{hh}^{(12)}(\bp,\bp)  &= &T(p;t_S)R(p;t_{M1})T(-p-\theta K_z+\rho q;t_A) R^*(p;t_S)R^*(p;t_{M2})R^*(p+\theta K_z-\rho q;t_A) \\
 & &\times \tj_{\rm in}(p,q,p,q+h\rho) \exp\big[ -\rmi [px_0 - qy_S + h(u+\theta\Delta z)] \big] \\
    \tj_{hh}^{(21)}(\bp,\bp)  &= &\tj_{hh}^{(12)*}(\bp,\bp) .
\end{eqnarray*}

\section{List of the main symbols}\label{appendix:6}

\begin{tabbing}
$\rb=(\bx,z)$ \hspace{15mm}        \= position vector \kill
$\hz$                              \> normal to the crystal surface \\
$\bx=(x,y)$                        \> $\rb$ component orthogonal to $\hz$  \\
$\bh=2\pi\hat{\mathbf{x}}\big/d$   \>reciprocal vector \\
$d$                                \>diffracting plane spacing \\
$\bK_o, \bK_h = \bK_o+\bh$         \>kinematical wave vectors \\
$2K\sin(\Theta_B)=h$      \>Bragg law \\
$\Theta_B$                \>Bragg angle \\
$\gamma = \cos(\Theta_B)$ \>$\bK_o$'s $z$ direction-cosine \\
$\alpha = \sin(\Theta_B)$ \>$\bK_h$'s $x$ direction-cosine \\
$K_z=K\gamma$             \>$z$ component of $\bK_{o,h}$ \\
$\bp=(p, q)$              \> variable conjugate to $\bx$ \\
$p$                       \>resonance error \\
$\chi_{0,h}$                       \>x-rays:  Fourier components of the \\ \>periodic electric susceptibility \\
$\upsilon_{0,h} = -K^2\chi_{0,h}$  \>neutrons: Fourier components of the \\ \>periodic Fermi pseudo-potential \\
$n_0 = 1+\Re(\chi_0)\big/2$        \>refractive index \\
$\mu_0 = \Im(\chi_0) K$            \>absorption coefficient \\
$\nu=\chi_h\big/|\chi_h|$ \\
$\Delta_e = \lambda\gamma\big/|\chi_h|$  \> pendell\"osung length \\
$\eta=\Delta_e \tan(\Theta_B) p\big/\pi$ \> dimensionless resonance error \\
$\zeta=\pi z\big/\Delta_e$               \> dimensionless propagation distance \\
$t_S, t_{M1}, t_{M2}, t_A$               \> crystal thicknesses \\
$z_A, z_D$                               \> analyser and detector distances\\ \>from the source\\
$x_0=x_1-x_2$           \> shear of the interfering beams \\
$\Delta z$              \> defocus \\
$y_S$                   \> start separation of the rays \\ \> ending collinearly \\
$y_D$                   \> end separation of the rays \\ \> starting collinearly \\
$u$                     \> analyser displacement along $\hat{x}$ \\
$\theta$                \> analyser yaw angle, rotation about $\hat{y}$ \\
$\rho$                  \> analyser pitch angle, rotation about $\hat{z}$ \\
$n=o, h$                \> particle state components (label) \\
$i=1, 2$                \> interferometer arm (label)
\end{tabbing}

\bibliography{INTn-3D}

\end{document}